\documentclass[aps,twocolumn,preprintnumbers,showpacs,showkeys
]{revtex4}
\usepackage{epsfig}
\usepackage{amssymb,amsmath,amsfonts,amsthm,graphicx
}
\graphicspath{{./Figures_col/}}  
\newcommand{\beq}{\begin{equation}}
\newcommand{\eeq}{\end{equation}}
\newcommand{\bea}{\begin{eqnarray}}
\newcommand{\eea}{\end{eqnarray}}
\newcommand{\Fig}[1]{Fig.~\ref{#1}}
\newcommand{\Tab}[1]{Table~\ref{#1}}
\newcommand{\Sec}[1]{Section~\ref{#1}}
\newcommand{\Eq}[1]{Eq.~(\ref{#1})}
\newcommand{\caa}{{\cal A}}

\newcommand{\tr}{\operatorname{Tr}}

\newcommand{\bc}{{\it bc~}}

\newcommand{\ageq}{\mbox{}_{\textstyle \sim}^{\textstyle > }}

\begin{document}
\preprint{HU-EP-13/07, ITEP-LAT/2013-1 -- revised version}

\title{Landau gauge ghost propagator and running coupling \\
in $SU(2)$ lattice gauge theory}

\author{V.~G.~Bornyakov}
\affiliation{Institute for High Energy Physics NRC "Kurchatov Institute", 
142281, Protvino, Russia \\
and Institute of Theoretical and Experimental Physics, 117259 Moscow, Russia\\
and School of Biomedicine, Far Eastern Federal University, 690950 
Vladivostok, Russia
}

\author{E.-M.~Ilgenfritz}
\affiliation{Joint Institute for Nuclear Research, BLTP, 141980
  Dubna, Russia} 

\author{C.~Litwinski}
\affiliation{Humboldt-Universit\"at zu Berlin, Institut f\"ur Physik, 
 12489 Berlin, Germany}

\author{V.~K.~Mitrjushkin}
\affiliation{Joint Institute for Nuclear Research, BLTP 141980 Dubna, Russia \\
and Institute of Theoretical and Experimental Physics, 117259 Moscow, Russia}

\author{M.~M\"uller--Preussker}
\affiliation{Humboldt-Universit\"at zu Berlin, Institut f\"ur Physik, 
  12489 Berlin, Germany}

\date{October 8, 2015}

\begin{abstract}
We study finite (physical) volume and scaling violation effects of the
Landau gauge ghost propagator as well as of the running coupling
$\alpha_s(p)$ in the $SU(2)$ lattice gauge theory. We consider lattices
with physical linear sizes between $aL \simeq 3$ and $aL \simeq 7$ fm and 
values of lattice spacing between $a=0.2$ and $a=0.07$ fm. 
To fix the gauge we apply an efficient gauge fixing method aimed at 
finding extrema as close as possible to the global maximum of the gauge 
functional. We find finite volume effects to be small for the 
lattice size $aL \simeq 3$ fm at momenta $|p|\, \ageq \, 0.6$ GeV. 
For the same lattice size we study extrapolations to the continuum limit 
of the ghost dressing function as well as for the running coupling with 
momenta chosen between $|p| = 0.41$ GeV and $|p| = 3.2$ GeV. We present fit 
formulae for the continuum limit of both observables in this momentum range.
Our results testify in favor of the decoupling behavior in the 
infrared limit. 
\end{abstract}



\maketitle

\section{Introduction}
\label{sec:introduction}

The infrared (IR) behavior of Landau gauge gluon and ghost propagators 
is believed to be closely related to gluon and quark confinement.
The celebrated Gribov--Zwanziger/Kugo--Ojima 
(GZKO) color confinement scenario ~\cite{Gribov:1977wm, Kugo:1979gm, Zwanziger:1991gz,
Zwanziger:2001kw, Zwanziger:2009je} has prescribed that the gluon propagator
$D(p)$ should vanish in the IR limit $p \to 0$ (the so-called infrared
suppression), while the ghost dressing function $p^2 G(p)$ was expected to 
become singular in this limit (infrared enhancement).

The search for gluon and ghost propagator solutions of Dyson-Schwinger (DS) 
and functional renormalization group (FRG) equations showed the existence 
of infrared solutions exhibiting a power-like {\it scaling} 
behavior~\cite{vonSmekal:1997is,
vonSmekal:1997vx, Alkofer:2000wg, Alkofer:2002aa, Fischer:2002hna,
Fischer:2002eq, Pawlowski:2003hq, Alkofer:2004it,Fischer:2006vf}.
Later also regular so-called {\it decoupling} solutions providing an
IR-finite limit of both the gluon propagator and the ghost dressing
function~\cite{Boucaud:2007va,Boucaud:2008ji,Boucaud:2008ky,Aguilar:2008xm,
Pennington:2011xs} have been found. Both kinds of solutions can be realized 
by different IR boundary conditions for the ghost dressing function as it 
has been argued in \cite{Fischer:2008uz}. 
As one understood immediately, both of them can support quark
confinement~\cite{Braun:2007bx}, and the gluon propagator breaks
reflection positivity. The decoupling solution 
cannot be reconciled with the GZKO scenario. However, it is in agreement 
with the refined Gribov-Zwanziger formalism developed in 
Refs.~\cite{Dudal:2007cw,Dudal:2008sp}.

From the phenomenological point of view the propagators can serve as input
to bound state equations as there are Bethe-Salpeter or Faddeev
equations for hadron phenomenology~\cite{Alkofer:2000wg,Alkofer:2006jf,
Eichmann:2008ef}. In the ultraviolet limit they allow a determination of
phenomenogically relevant parameters such as $~\Lambda_{\overline{MS}}~$
or condensates $\langle \overline{\psi}\psi \rangle,~\langle A^2 \rangle,
\ldots$, by fitting ab-initio lattice data to continuum 
expressions (see e.g. ~\cite{Sternbeck:2012qs,Burger:2012ti} and references
therein) obtained from operator product expansion and perturbation
theory~\cite{Chetyrkin:2004mf, Chetyrkin:2009kh}.

What concerns the solution of DS and/or FRG equations it is well-known
that in practice the system of those equations is truncated. 
The details of truncation influence the behavior of the Green functions
especially in the non-perturbative momentum range 
around $~1 \,\mathrm{GeV}$, where the Landau gauge gluon dressing function 
exhibits a pronounced maximum. Therefore, reliable results from ab-initio 
lattice computations to compare with or even used as an input for DS or FRG
equations are highly welcome.

On the lattice, over almost twenty years extensive studies of the
Landau gauge gluon and - in the present paper discussed again - 
ghost propagators have been carried out (see, e.g.,
\cite{Mandula:1987rh,Suman:1995zg,Leinweber:1998uu,Becirevic:1999uc,
Becirevic:1999hj,Bonnet:2000kw,Bonnet:2001uh,Bloch:2002we,
Bloch:2003sk,Furui:2003jr,Boucaud:2005gg,
Sternbeck:2005tk,Bowman:2007du,Cucchieri:2007md,Cucchieri:2007rg,
Cucchieri:2007zm,Sternbeck:2007ug,Cucchieri:2008fc,Oliveira:2008uf,
Bogolubsky:2009dc,Oliveira:2012eh, Maas:2014xma}).  
A serious problem in these calculations represents the ambiguity  
of Landau gauge fixing (the Gribov copy problem) 
\cite{Cucchieri:1997dx,Bakeev:2003rr,Furui:2003mz,Silva:2004bv,Furui:2004cx,
Bogolubsky:2005wf,Bornyakov:2008yx,Maas:2008ri,Maas:2009ph,Hughes:2012hg}.
As long as the latter is solved by extremizing the Landau gauge
functional (for alternative approaches see 
\cite{Maas:2009se,Sternbeck:2012mf})
numerical lattice results clearly support the decoupling-type
of solutions in the IR limit and the lack of IR enhancement of
the ghost propagator ~\cite{Cucchieri:2007md,Cucchieri:2007rg,
Cucchieri:2008fc,Bornyakov:2008yx,Bogolubsky:2009dc,Maas:2014xma}. 
For the gluon propagator D. Zwanziger recently has derived 
a strict bound $\lim_{p \to 0} ~p^{d-2} D(p)=0$ also allowing 
$D(0) \ne 0$ for $d>2$~\cite{Zwanziger:2012xg}, i.e. a decoupling
behavior (see also \cite{Cucchieri:2012cb}). Note, that for such
a behavior it became more and more evident that BRST symmetry is 
broken \cite{vonSmekal:2008es,Burgio:2009xp,Cucchieri:2014via,
Cucchieri:2014xfa}.

However, most of the lattice computations dealing 
with the IR limit were relying on rather coarse lattices 
in order to reach large enough volumes. A systematic
investigation of lattice discretization artifacts or scaling violations
and an extrapolation to the continuum was missing for quite a long time.

In this paper we present an investigation for the ghost dressing
function and -- employing previous gluon propagator results
\cite{Bornyakov:2009ug} -- obtain the running coupling within the so-called
minimal MOM scheme \cite{vonSmekal:2009ae}.
We restrict ourselves to the $SU(2)$ case of pure gauge theories,
having in mind the close similarity to the more realistic $SU(3)$ 
case as observed in \cite{Cucchieri:2007zm,Sternbeck:2007ug}.

We shall use the same lattice field configurations as in 
\cite{Bornyakov:2009ug} which were gauge fixed with an improved method 
taking into account many copies over all $Z(2)$ Polyakov loop sectors and 
applying simulated annealing with subsequent overrelaxation. 
We separately discuss the case of fixed lattice spacing and 
varying volume (from $aL \simeq 3$~fm to $aL \simeq 7$~fm) and the 
case of fixed physical volume and varying lattice spacing 
(between $a=0.21$~fm and $a=0.07$~fm). In the range of IR momenta 
achieved in this setting, finite-size effects are
shown to be negligibly small. But relative finite-discretization 
effects in the infrared (for a renormalization scale chosen at 
$\mu=2.2$~GeV) turn out to be more sizable and can be quantified to 
reach a 10 percent variation level at $p\simeq 0.4$~GeV in the 
approximate scaling region explored between 
$\beta=2.3$ and $\beta=2.55$. 
A similar observation can be made for the running 
coupling. Therefore, a careful analysis of the lattice artifacts 
is mandatory. We carry out such an analysis by taking the continuum
limit extrapolations (for the first time to our knowledge)
for the ghost propagator as well as for the running coupling  
for selected physical momentum values in the range
from $|p|=0.41$ GeV to $|p|=3.2$ GeV. The continuum extrapolated values can
then be fitted with appropriate formulae describing a smooth continuum
behavior of the observables in the given momentum range.   

In \Sec{sec:definitions} we introduce the lattice Landau gauge 
and the corresponding Faddeev-Popov operator and the ghost propagator.  
In \Sec{sec:details} some details of the simulation and of the improved 
gauge fixing are repeatedly given for the convenience of the reader.
In \Sec{sec:results} we present our numerical results for the ghost
propagator and the running coupling. Conclusions will be drawn in 
\Sec{sec:conclusions}.

\section{Lattice Landau gauge and the ghost propagator}
\label{sec:definitions}

Let us briefly recall how the $SU(2)$ gauge field configurations  
used in Ref. \cite{Bornyakov:2009ug} for measuring the gluon propagator
have been created and gauge fixed. 

The non-gauge-fixed $SU(2)$ gauge field configurations were generated
with a standard Monte Carlo routine using the standard plaquette 
Wilson action  
\bea
S &=& \beta {\sum}_{x} {\sum}_{\mu >\nu}
\left[ 1 -\frac{1}{2}~\tr \Bigl(U_{x\mu}U_{x+\hat{\mu};\nu}
U_{x+\hat{\nu};\mu}^{\dagger}U_{x\nu}^{\dagger} \Bigr)\right],\nonumber \\
 \beta &=& 4/g_0^2 ,
\label{eq:action}
\eea
where $g_0$ denotes the bare coupling constant. The link variables 
$U_{x\mu} \in SU(2)$ 
transform under local gauge transformations $g_x$ as follows 
\beq
U_{x\mu} \stackrel{g}{\mapsto} U_{x\mu}^{g}
= g_x^{\dagger} U_{x\mu} g_{x+\hat{\mu}} \,,
\qquad g_x \in SU(2) \,.
\label{eq:gaugetrafo}
\eeq
The standard (linear) definition~\cite{Mandula:1987rh} for the 
dimensionless lattice gauge vector potential $\caa_{x+\hat{\mu}/2,\mu}$ 
is
\beq
\caa_{x+\hat{\mu}/2,\mu} = \frac{1}{2i}~\Bigl( U_{x\mu}-U_{x\mu}^{\dagger}\Bigr)
\equiv A_{x+\hat{\mu}/2;\mu}^a \frac{\sigma_a}{2} \,.
\label{eq:a_field}
\eeq
The definition of the gluon field is not unique at finite $a$, which may 
influence the propagator results in the IR region, where the continuum limit 
is more difficult to control.

In lattice gauge theory the most natural choice of the Landau gauge 
condition is by transversality~\cite{Mandula:1987rh}
\beq
(\partial \caa)_{x} = {\sum}_{\mu=1}^4 \left( \caa_{x+\hat{\mu}/2;\mu}
  - \caa_{x-\hat{\mu}/2;\mu} \right)  = 0 \,,
\label{eq:diff_gaugecondition}
\eeq
which is equivalent to finding a local extremum of the gauge functional
\beq
F_U(g) = ~\frac{1}{4V}{\sum}_{x\mu}~\frac{1}{2}~\tr~U^{g}_{x\mu} 
\label{eq:gaugefunctional}
\eeq
with respect to gauge transformations $~g_x~$. $V=L^4~$ denotes the $4d$
lattice size.The Gribov ambiguity is reflected by the existence of 
multiple local extrema. The manifold consisting of Gribov copies providing 
local maxima of the functional (\ref{eq:gaugefunctional}) and a semi-positive 
Faddeev-Popov operator (see below) is called the {\it Gribov region} $~\Omega$,
while the global maxima form what is called the {\it fundamental modular 
region} (FMR) $~\Lambda \subset \Omega$.  Our gauge fixing procedure 
is aiming to approach $~\Lambda$ by finding higher and higher maxima.
This is achieved by use of the effective optimization algorithm and 
finding a large number of local maxima of which the highest is picked up.
\par
The lattice expression of the Faddeev-Popov operator $M^{ab}$
corresponding to $M^{ab} = - \partial_{\mu} D^{ab}_{\mu}$ in the
continuum theory (where $D^{ab}_{\mu}$ is the covariant derivative 
in the adjoint representation) is given by
\cite{Zwanziger:1993dh,Suman:1995zg}
\bea
M^{ab}_{xy}  =  {\sum}_{\mu}~\Bigl\{
  \left( \bar{S}^{ab}_{x\mu} + \bar{S}^{ab}_{x-\hat{\mu};\mu}
\right)~\delta_{x;y} \Bigr. \nonumber     \\
\Bigl.   - \left( \bar{S}^{ab}_{x\mu} - \bar{A}^{ab}_{x\mu}
\right)~\delta_{y;x+\hat{\mu}} \Bigr.  \\
\Bigl.   - \left( \bar{S}^{ab}_{x-\hat{\mu};\mu}
+ \bar{A}^{ab}_{x-\hat{\mu};\mu} \right)~\delta_{y;x-\hat{\mu}}
\Bigr\} \nonumber
\label{eq:M-form3}
\eea
where
\beq
\bar{S}^{ab}_{x\mu} = \delta^{ab}~\frac{1}{2}~\tr~U_{x\mu},~~
\bar{A}^{ab}_{x\mu} = -\frac{1}{2}~\epsilon^{abc}~A_{x+\hat{\mu}/2;\mu}^c .
\label{eq:abbreviations}
\eeq
From the form (\ref{eq:M-form3}) it follows that a trivial zero eigenvalue 
is always present, such that at the Gribov horizon $\partial \Gamma$ the 
first  non-trivial zero eigenvalue appears.  
For configurations with a constant field, 
with $b^{0}_{x\mu} = \bar{b}^{0}_{\mu}$ and
$b^{a}_{x\mu}=\bar{b}^{a}_{\mu}$ independent of $x$, there exist
eigenmodes of $M$ with a vanishing eigenvalue.
Thus, if the Landau gauge is properly implemented,
$M[U]$ is a symmetric and semi-positive definite matrix.

The ghost propagator $G^{ab}(x,y)$ is defined
as~\cite{Zwanziger:1993dh,Suman:1995zg}
\beq
G^{ab}(x,y) = \delta^{ab}~G(x-y) \equiv
\Bigl<\,\left(\, M^{- 1}\,\right)^{a\,b}_{x\, y} [U]\,\Bigl> \; ,
\label{ghostprop}
\eeq
where $M[U]$ is the Faddeev-Popov operator, on the sector orthogonal to
the strict zero modes. Note that the ghost propagator becomes translationally
invariant ({\it i.e.}, dependent only on $x-y$) and diagonal in color space 
only in the result of averaging over the ensemble of gauge-fixed representatives
of the original gauge-unfixed Monte Carlo gauge ensemble. 

$M[U]$ can be inverted with a conjugate-gradient method, provided
that both the source $\psi^{a}(y)$ and the initial guess for the solution
are orthogonal to the zero modes. For the source we adopt the one proposed 
in \cite{Cucchieri:1997dx} and also used in \cite{Bakeev:2003rr}:
\beq
\psi^{a}(y) \,=\,\delta^{a c} \, e^{ 2 \pi i \, p \cdot y}
\qquad p \neq (0,0,0,0) \; ,
\label{eq:source}
\eeq
for which the condition ${\sum}_{y}\, \psi^{a}(y)\,=\,0$ is
automatically imposed. Only the scalar product of $M^{-1}\psi$
with the source $\psi$ itself has to be evaluated.
The inversion of $M$ is done on sources for fixed $c = 1,..,3$ and 
the (adjoint) color averaging will be explicitely performed.

The ghost propagator in momentum space can be written as
\beq
G(p)\, = \,\frac{1}{3 V} {\sum}_{x\mbox{,}\, y} e^{- 2 \pi i \, p \cdot (x - y)}
\Bigl<\,\left(\, M^{- 1}\,\right)^{a\,a}_{x\, y} [U]\,\Bigl> \; ,
\label{eq:ghostprop_in_momentumspace}
\eeq
where the coefficient $\frac{1}{3V}$ is taken for a full normalization, 
including the indicated color average over $a=1,..,3$.
In what follows we will denote the (bare) ghost dressing function as
\beq
J(p)\equiv p^2 G(p)\;.
\label{eq:ghostdressing_fct}
\eeq

\section{Details of the computation}
\label{sec:details}

The Monte Carlo (MC) simulations had been carried out at several 
$\beta$-values between $\beta=2.2$ and $\beta=2.55$ for various 
lattice sizes $L$. Consecutive configurations (considered to be 
statistically independent) were separated by 100 sweeps, each sweep 
consisting of one local heatbath update followed by $L/2$ microcanonical 
updates. In \Tab{tab:data_sets} we provide the full information 
about the field ensembles used in this investigation. 
The corresponding results concerning the gluon propagator
have been published in \cite{Bornyakov:2009ug}.
%
\begin{table}[h]
\begin{center}
\vspace*{0.2cm}
\begin{tabular}{|c|c|c|c|c|c|c|} \hline
$\beta$ & $a^{-1}$ [GeV] & $a$ [fm] & $~L~$ & $aL$ [fm] & 
$~N_{conf}~$ & $N_{copy}$ \\ \hline\hline

 2.20  & 0.938 & 0.210 &  14  & 2.94  & 400  &  48     \\ 
 2.30  & 1.192 & 0.165 &  18  & 2.97  & 200  &  48     \\
 2.40  & 1.654 & 0.119 &  26  & 3.09  & 200  &  48     \\ 
 2.50  & 2.310 & 0.085 &  36  & 3.06  & 400  &  80     \\ 
 2.55  & 2.767 & 0.071 &  42  & 2.98  & 200  &  80     \\ \hline\hline
                                         
 2.20  & 0.938 & 0.210 &  24  & 5.04  & 400  &  48     \\ 
 2.30  & 1.192 & 0.165 &  30  & 4.95  & 400  &  48     \\ 
 2.40  & 1.654 & 0.119 &  42  & 5.00  & 200  &  80     \\ \hline\hline

 2.30  & 1.192 & 0.165 &  44  & 7.26  & 200  &  80     \\ \hline\hline

\end{tabular}
\end{center}
\caption{Values of $\beta$, lattice sizes, number of configurations
and number of gauge copies used throughout Ref. \cite{Bornyakov:2009ug}
and this paper. The lattice spacing was fixed to its physical value
using the string tension $\sqrt{\sigma}=440$ MeV (see 
\cite{Fingberg:1992ju,Lucini:2001ej}).
} 
\label{tab:data_sets}
\end{table}
 
The gauge fixing is completed by the $Z(2)$ flip operation as discussed in
\cite{Bogolubsky:2005wf,Bogolubsky:2007bw}. For the convenience of the reader
we briefly recall the main features. The method consists in flipping all link 
variables $U_{x\mu}$ attached and orthogonal to a selected $3d$ plane by 
multiplying them with $-1 \in Z(2)$. Such global flips are equivalent 
to non-periodic gauge transformations. They represent an exact 
symmetry of the pure gauge action. The Polyakov loops in the direction of 
the chosen links and averaged over the orthogonal $3d$ plane (base space)
obviously change their sign. Therefore, the flip operations combine the 
$2^4$ distinct gauge orbits (or Polyakov loop sectors) related to strictly 
periodic gauge transformations into a single large gauge orbit.

The second ingredient is the simulated annealing (SA) method, 
which has been investigated independently and found computationally 
more efficient than the exclusive use of standard overrelaxation 
(OR) \cite{Schemel:2006da,Bogolubsky:2007pq,Bogolubsky:2007bw}.
The SA algorithm generates gauge transformations $~g(x)~$ by MC 
iterations with a statistical weight proportional to
$~\exp{(4V~F_U[g]/T)}~$. The ``gauge temperature'' $~T~$ is an auxiliary
parameter which is gradually decreased (during gauge fixing a configuration)
in order to guide the gauge functional $~F_U[g]~$ towards a maximum, 
despite its fluctuations.  
In the beginning, $~T~$ has to be chosen sufficiently large in order 
to allow rapidly traversing the configuration space of $~g(x)~$ 
fields in large steps. As in Ref. \cite{Bogolubsky:2007bw} we have 
chosen $~T_{init}=1.5$. After each quasi-equilibrium sweep (that
includes both heatbath and microcanonical updates) $~T~$ has been 
decreased in equidistant steps. The final SA temperature has been 
chosen according to the requirement that during the subsequent 
execution of the OR algorithm the violation of the transversality 
condition
\beq
\max_{x\mbox{,}\, a} \, \Big|
{\sum}_{\mu=1}^4 \left( A_{x+\hat{\mu}/2;\mu}^a - A_{x-\hat{\mu}/2;\mu}^a \right)
\Big| \, < \, \epsilon_{lor}
\label{eq:gaugefixstop}
\eeq
decreases in a {\it monotonous manner} for the majority of gauge fixing 
trials, until finally the transversality condition (\ref{eq:gaugefixstop}) 
becomes uniformly satisfied with an $\epsilon_{lor}=10^{-7}$.
Such a monotonous OR behavior is reasonably satisfied for a lower 
gauge temperature value $~T_{final}=0.01~$~\cite{Schemel:2006da}, to be 
reached in the last step of SA.  
The number of temperature steps of SA interpolating between $~T_{init}$ and
$~T_{final}$ has been chosen to be $1000$ for the smaller lattice sizes 
and has been increased to $2000$ for the lattice sizes $30^4$ and bigger.
The finalizing OR algorithm using the Los Alamos type overrelaxation
with the overrelaxation parameter value $\omega = 1.7$ requires typically 
a number of iterations varying from $O(10^2)$ to $O(10^3)$ before the 
configuration can be considered as gauge-fixed with the above mentioned
precision $\epsilon_{lor}$.

In what follows we call the combined algorithm employing SA (with 
finalizing OR) and $Z(2)$ flips the `FSA' algorithm. By repeated starts 
of the FSA algorithm we explore each $Z(2)$ Polyakov loop sector 
{\it several times} in order to find there the best (``\bc'') copy
\footnote{An alternative idea to tackle the Gribov problem has been 
discussed in \cite{Maas:2009se,Sternbeck:2012mf}.}.
The total number of copies per configuration $N_{copy}$
for each $\beta$-value and lattice size, generated and inspected for 
selecting the optimal $F_U(g)$, 
is indicated in \Tab{tab:data_sets}.

Some more details suitable to speed up the gauge fixing procedure are 
described in \cite{Bornyakov:2008yx}.

In order to suppress lattice artifacts in the propagators 
we followed Ref.~\cite{Leinweber:1998uu} and 
selected the allowed lattice momenta as surviving the {\it cylinder cut}  
\beq
{\sum}_\mu k^2_\mu - \frac{1}{4}({\sum}_\mu k_\mu )^2 \leq 1 \,.
\eeq
Moreover, we have applied the  ``$\alpha$-cut'' \cite{Nakagawa:2009zf} 
$p_{\mu} \le (2/a) \alpha$ for every component, in order to
keep close to a linear behavior of the lattice momenta 
$p_{\mu}=(2 \pi k_{\mu})/(aL), ~~k_{\mu} \in (-L/2,L/2]$. 
We have chosen $\alpha=0.5$. Obviously, this cut influences large 
momenta only.
 
We define the renormalized ghost dressing function
according to momentum subtraction schemes (MOM) by
\bea
J_{ren}(p,~\mu) ~&=& {\cal Z}(\mu,1/a) ~J(p, 1/a)\,, \\
J_{ren}(p=\mu) &=& 1.
\label{eq:renorm}
\eea
In practice, we have fitted the bare dressing function 
$J(p, 1/a)$ with an appropriate function (see \Eq{eq:fitghost} below) 
and then used the fits for renormalizing $J$. Assuming that lattice
artifacts are sufficiently suppressed it has to be seen, whether 
multiplicative renormalizability really holds in the non-perturbative 
regime. 
For this it is sufficient to prove that ratios of the renormalized 
(or  unrenormalized) propagators obtained from different cutoff values  
$1/a(\beta)$ will not depend on $p$ 
at least within a certain  momentum interval $[p_{min}, p_{max}]$, 
where $p_{max}$ should be the maximal momentum surviving all the cuts 
applied.  

In what follows the subtraction momentum has been chosen as
$\mu=2.2$~GeV. 

\section{Results}
\label{sec:results}

\subsection{Ghost dressing function}

\begin{figure*}[tbh]
\centering
\includegraphics[width=5.9cm,angle=270]{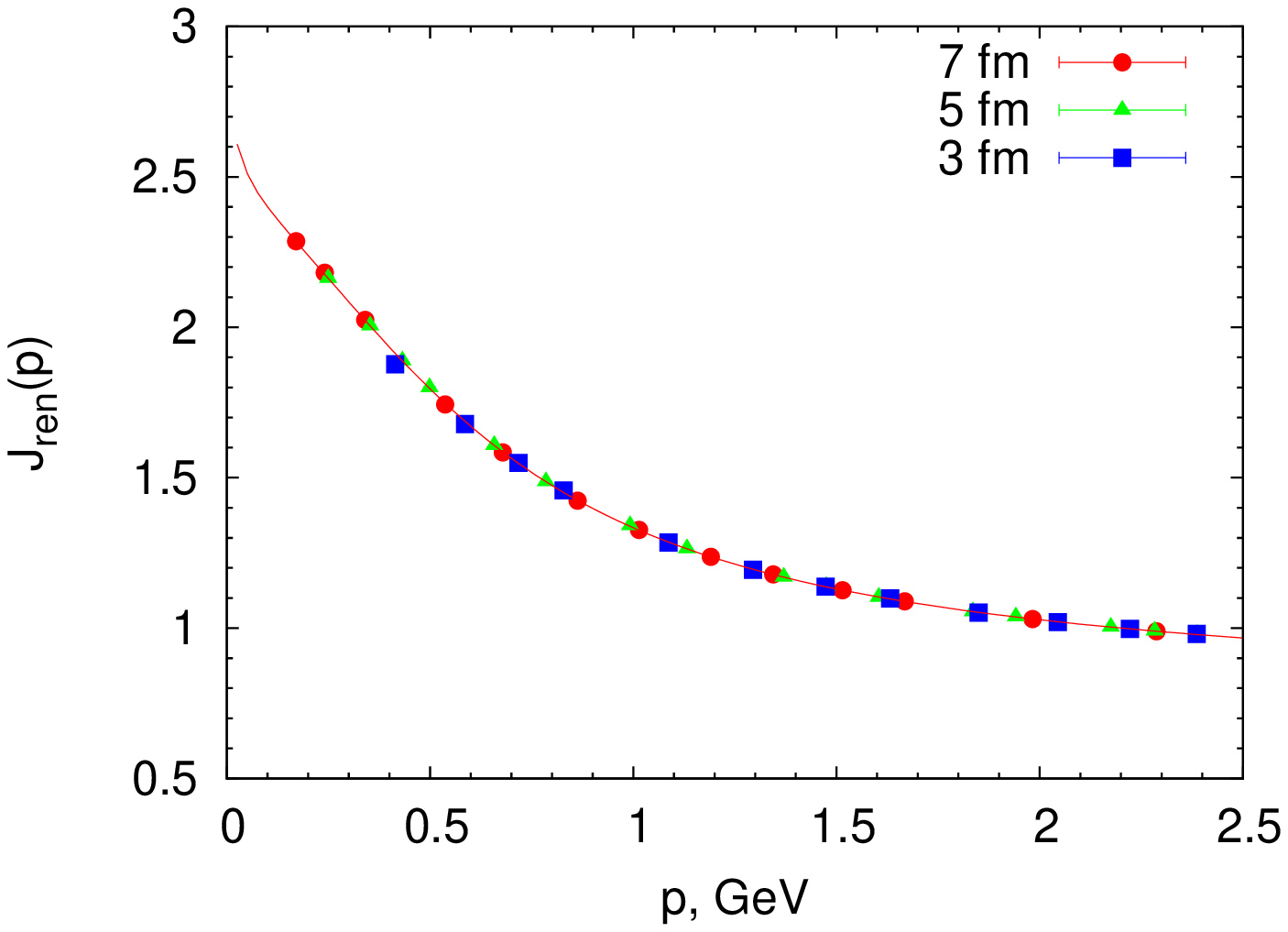}
\includegraphics[width=5.9cm,angle=270]{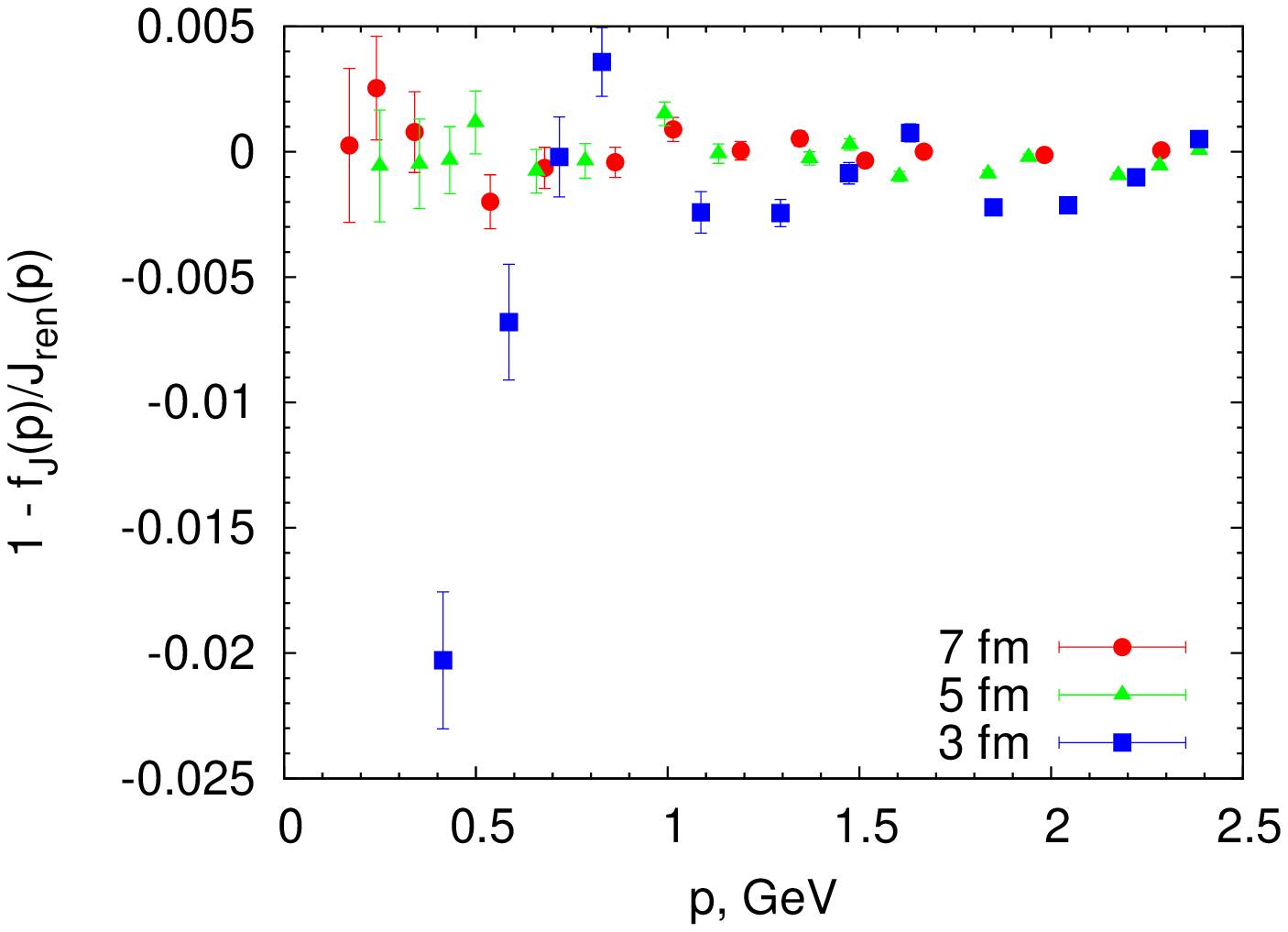}
\caption{{\bf Left:} The momentum dependence of the renormalized ghost dressing 
function $J_{ren}(p)$ for three different lattice sizes at $\beta=2.3$. 
The curve shows the fit applying \Eq{eq:fitghost} to the case $aL \simeq 7$~fm. 
{\bf Right:} The relative deviation of the data for the ghost dressing function 
$J_{ren}(p)$ from the applied fitting curve.
}
\label{fig:gdf_ren_2p3}
\end{figure*}
\begin{figure*}[tbh]
\centering
\includegraphics[width=5.9cm,angle=270]{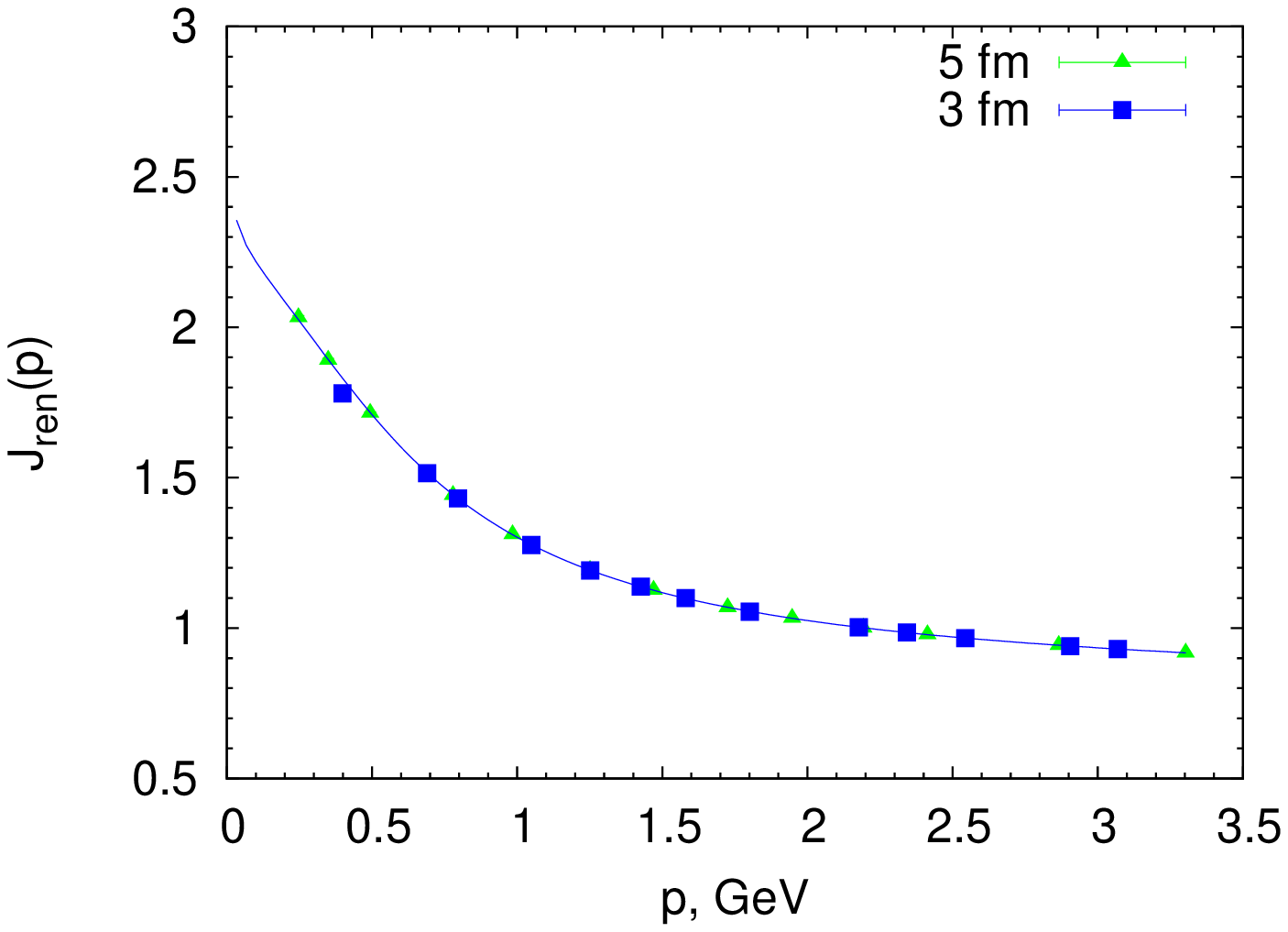}
\includegraphics[width=5.9cm,angle=270]{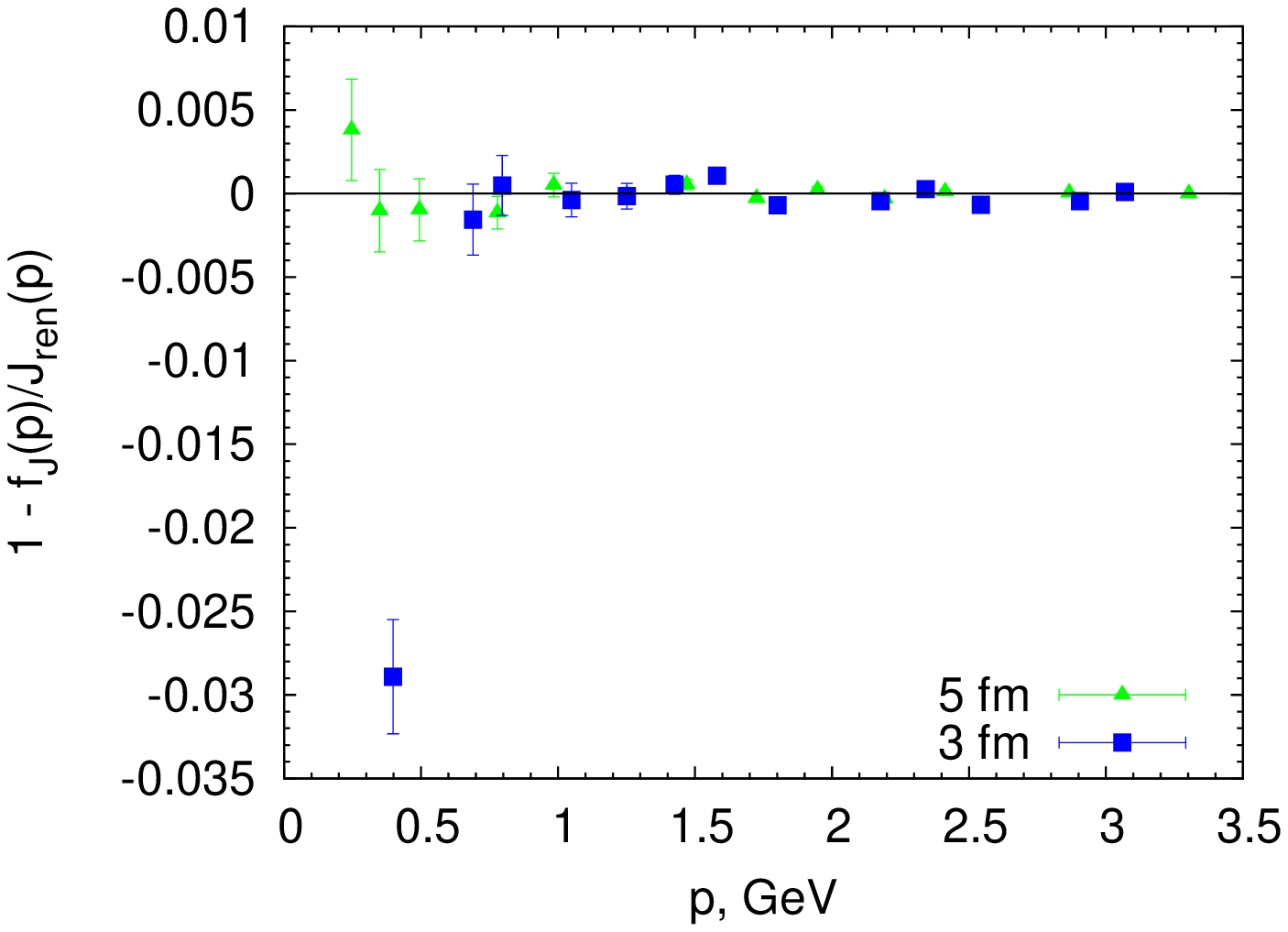}
\caption{The same as in \Fig{fig:gdf_ren_2p3} but for two lattice sizes 
at $\beta=2.4$. The curve shows the fit applying \Eq{eq:fitghost} to the 
case $aL \simeq 5$~fm.
}
\label{fig:gdf_ren_2p4}
\end{figure*}

First let us discuss the finite volume effects for the renormalized 
ghost dressing function $J_{ren}(p,\mu)$. The data for various volumes 
are presented in \Fig{fig:gdf_ren_2p3} 
for $\beta=2.3$ and in \Fig{fig:gdf_ren_2p4}  for $\beta=2.4$. 
To present the finite volume effects in more detail
we fitted the data at $\beta=2.3$ for $aL \simeq 7$~fm and at $\beta=2.4$ for 
$aL \simeq 5$~fm with a fitting function of the form
\beq
f_J(p) = \frac{b_1}{\hat{p}^{2\kappa}} +  \frac{b_2 \hat{p}^2}{1 + \hat{p}^2}
\label{eq:fitghost}
\eeq
with the dimensionless rescaled momentum $\hat{p} \equiv p/m_{gh}$
(see \Tab{tab:fit_ghost}).
This ansatz, while describing the data reasonably well within the given
momentum range, will not be applicable in the IR limit,
when we assume that $J(p)$ exhibits an inflection point and
bends to a finite value $J(0)$.

In the right panels of \Fig{fig:gdf_ren_2p3} and \Fig{fig:gdf_ren_2p4},
respectively, the relative deviations from the fit function
are shown for $\beta=2.3$ and $\beta=2.4$, respectively.
One can see that for both $\beta$ values finite volume effects for 
lattices even with $aL \simeq 3$~fm are small (less than 1\%) for momenta 
$|p|\, \ageq \, 0.6$ GeV.

\begin{figure*}[tb]
\centering
\includegraphics[width=5.9cm,angle=270]{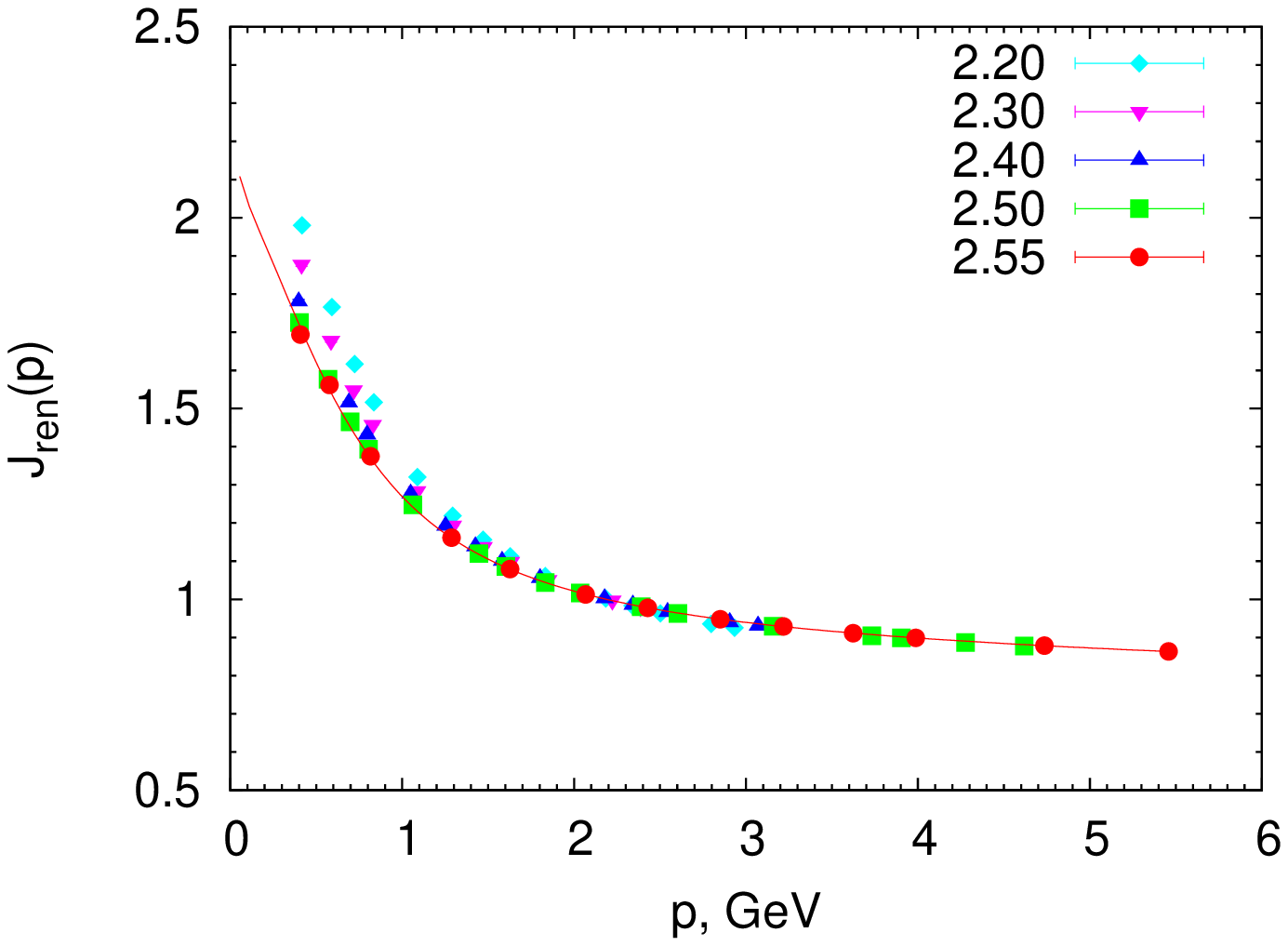}
\includegraphics[width=5.9cm,angle=270]{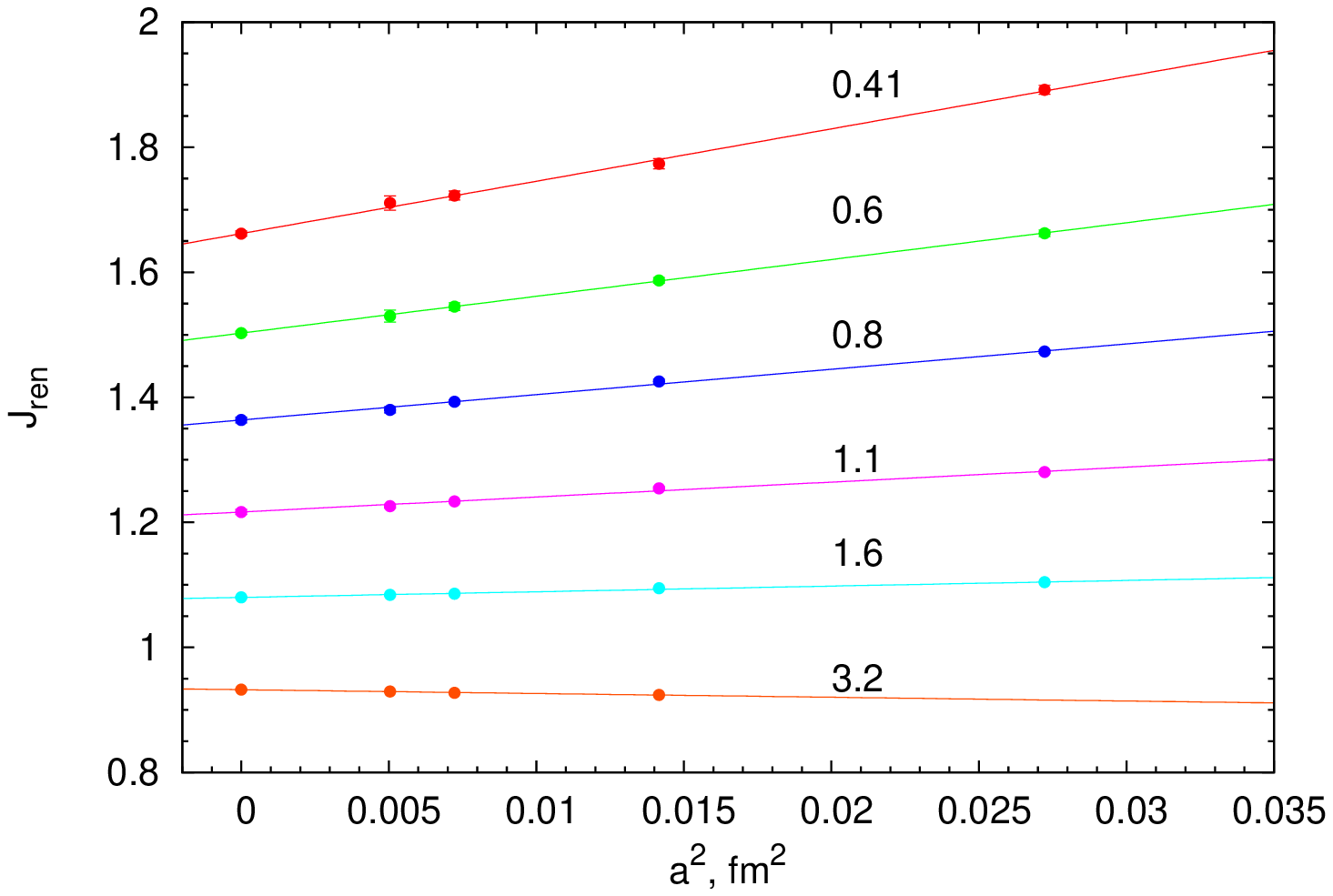}
\caption{{\bf Left:} The momentum dependence of the renormalized 
ghost dressing function $J_{ren}(p)$ for five different $\beta$-values or
lattice spacings. The physical linear box size is $aL \simeq 3$~fm. The fitting 
curve belongs to the smallest available lattice spacing ($\beta=2.55$). 
{\bf Right:} The continuum limit extrapolation for the ghost 
dressing function $J_{ren}(p)$. The lines show fits linear in $a^2$.
The numbers indicate the momentum values $p$ in GeV. 
Corresponding points related to $\beta=2.2$ are not taken 
into account for the linear fit and, thus, not shown.
The left most data points show the continuum extrapolated results.
}
\label{fig:gdf_ren_3fm}
\end{figure*}

Now let us come to the discussion of lattice artifacts.
In \Fig{fig:gdf_ren_3fm} (left) we show the momentum dependence of the 
renormalized ghost dressing function $J_{ren}(p)$ for five different
lattice spacings but for (approximately) the same physical size 
$aL \simeq 3$~fm  (for the exact values see \Tab{tab:data_sets}). 
Finite-spacing effects for $\beta=2.2, 2.3, 2.4$ in 
comparison with $\beta=2.55$ are evident.
The curve shows the fit function \Eq{eq:fitghost} for $\beta=2.55$ 
(see \Tab{tab:fit_ghost}).

For every $\beta$ value we computed the ghost propagator for chosen values of 
the momentum in the range $0.41$ GeV $\le p \le 3.2$ GeV  by interpolating the
data using the function \Eq{eq:fitghost}. For the interpolation 4 or 5 adjacent 
data points were used. For the purpose of this interpolation the choice of the function
\Eq{eq:fitghost} was not really important. We then computed the ghost propagator
in the continuum limit for these values of the momentum using a linear in $a^2$ 
extrapolation as shown in the right panel of \Fig{fig:gdf_ren_3fm}.
The data for the (comparably strong) coupling value $\beta=2.2$ were not 
used for the extrapolation and are not shown in this figure. 

\begin{figure*}[htb]
\centering
\includegraphics[width=5.9cm,angle=270]{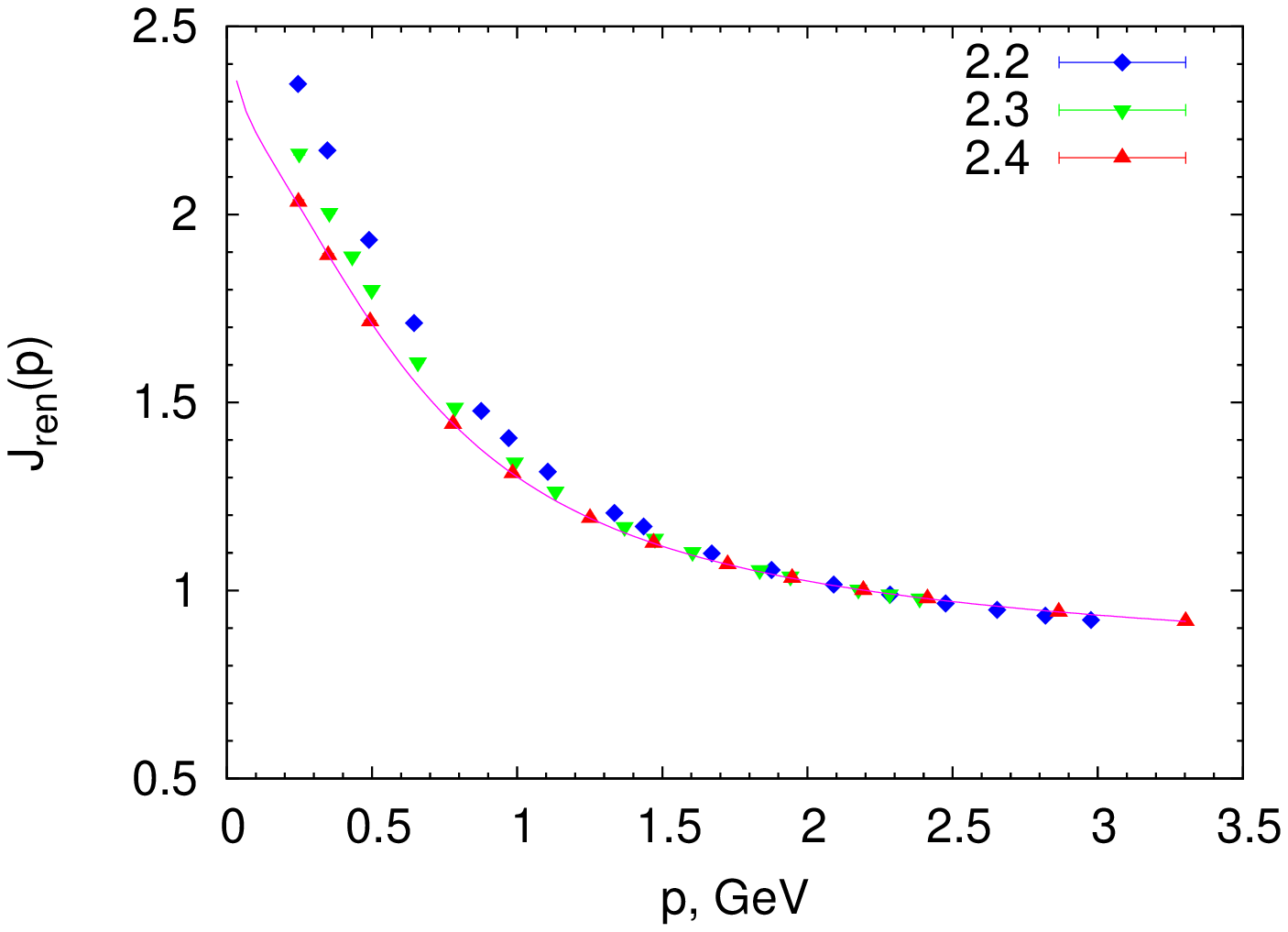}
\includegraphics[width=5.9cm,angle=270]{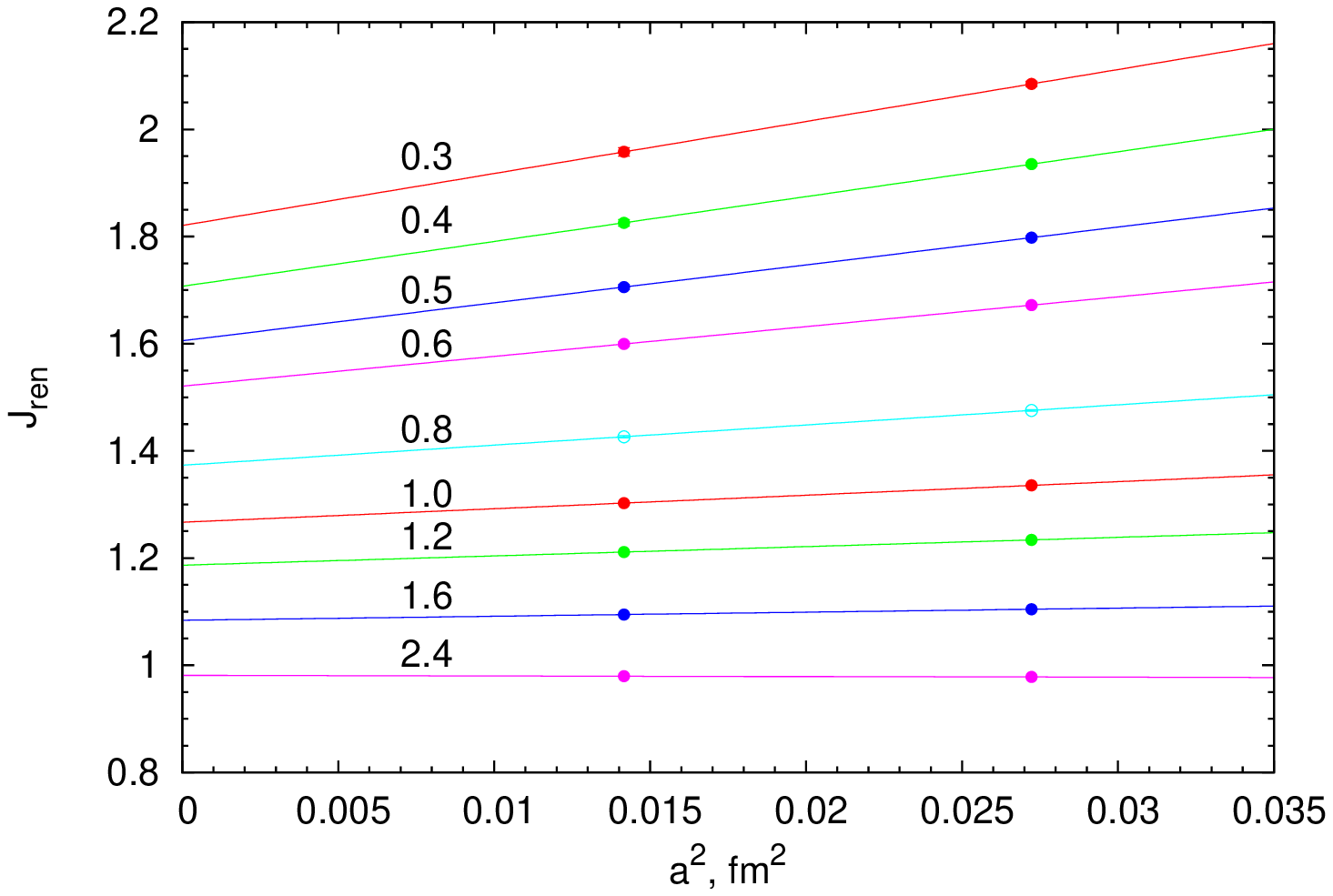}
\caption{{\bf Left:} Same as in \Fig{fig:gdf_ren_3fm} but for a linear box size of
$aL \simeq 5$~fm and three different $\beta$-values. 
The fitting curve belongs to $\beta=2.40$.
{\bf Right:} Dressing function $J_{ren}(p)$ for few selected momenta as function 
of $a^2$. 
Data points related to the stronger coupling value $\beta=2.2$ are again discarded.
The straight lines are only to guide the eye.
}
\label{fig:gdf_ren_5fm}
\end{figure*}

Related to our choice of the (re)normalization momentum $\mu=2.2$~GeV and
due to the rather small statistical errors for the ghost dressing function
we see clear scaling violations especially in the IR region but also for 
$p > \mu$.
At the lowest (here accessible) momenta the violations at 
$\beta=2.3$  ($\beta=2.55$) relative to the continuum limit value 
are staying below $14\%$ ($3\%$). 

Thus, in comparison with corresponding estimates for the 
gluon propagator (see Fig. 13 in \cite{Bornyakov:2009ug}) which were 
more noisy, we can say that the relative scaling violations of the ghost 
dressing function turn out to be somewhat larger.

Similar to the case $aL \simeq 3$~fm we observe analogous lattice spacing 
effects on volumes with linear size $aL \simeq 5$~fm. The respective 
results are depicted in \Fig{fig:gdf_ren_5fm}. 
As for the smaller volume we discard the data for $\beta=2.2$.
Under these circumstances a real extrapolation to the continuum 
limit cannot be done. Nevertheless, \Fig{fig:gdf_ren_5fm} (right) 
clearly demonstrates finite lattice effects of a strength similar 
to the smaller volume case.

\begin{table}[h]
\begin{center}
\vspace*{0.2cm}
\begin{tabular}{|c|c|c|c|c|c|c|} \hline 
$\beta$ & $L$ & $m_{gh}~[GeV]$ & $\kappa$ & $b_1$ & $b_2$ & $\chi^2_{df}$ \\ 
\hline
 2.30  & 44 & 0.64(1)   & 0.026(1)  & 2.20(1) & -1.15(1) &  1.8\\
\hline
 2.40  & 42 & 0.67(1)   & 0.0232(4) & 2.05(1) & -1.03(1) & 1.9 \\
\hline
 2.55  & 42 & 0.696(15) & 0.0215(4) & 1.90(2) & -0.89(2)&  2.27 \\
\hline\hline
 c.l.1 &    & 0.743(5)  & 0.016(1)  & 1.827(5) & -0.85(1) & 0.07\\
\hline
 c.l.2 &    & 0.708(5)  & 0.027(1)  & 1.898(5) & -0.930(4)& 0.04\\
\hline
\end{tabular}
\end{center}
\caption{Values of the fit parameters according to \Eq{eq:fitghost} 
and the corresponding $\chi^2_{df}$.
The last two lines correspond to fits of the extrapolated 
continuum limit values of the renormalized ghost dressing
function in accordance with \Eq{eq:fitghost} and \Eq{eq:fitghost2},
respectively, for lattice size $aL \simeq 3$ fm.
} 
\label{tab:fit_ghost}
\end{table}

\begin{figure}[tb]
\centering
\includegraphics[width=5.9cm,angle=270]{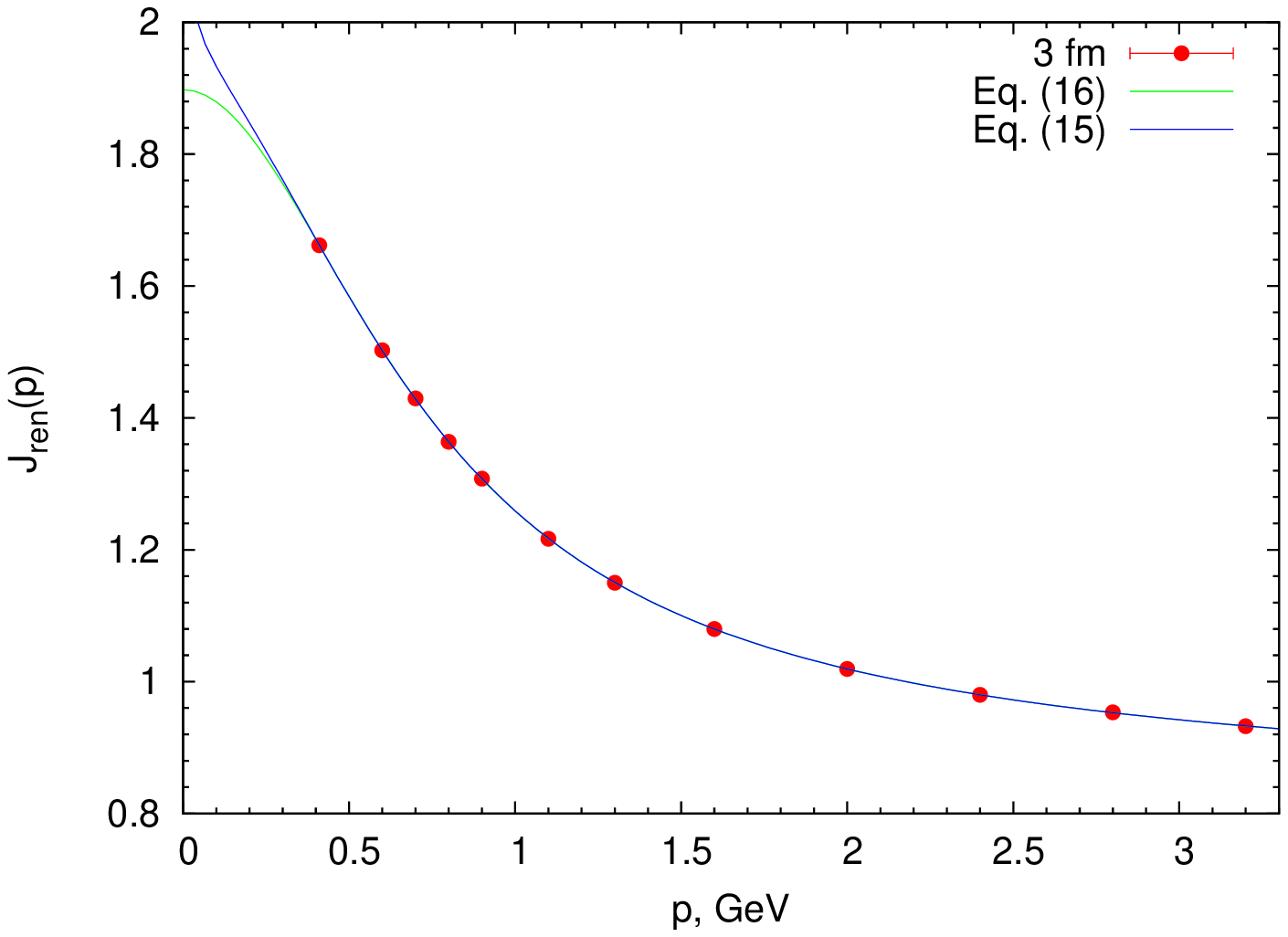}
\caption{The momentum dependence of the renormalized ghost dressing function $J_{ren}(p)$ 
extracted in the continuum limit for selected momenta for $aL \simeq 3$~fm. 
The curves show fits with the ansatzes \Eq{eq:fitghost} and \Eq{eq:fitghost2},
respectively.
}
\label{fig:Jpcont}
\end{figure}

Finally, let us present the continuum extrapolated result 
for the smaller volume of $aL \simeq 3$ fm in \Fig{fig:Jpcont}.
We show the extracted points together with two fit curves: one with
the ansatz \Eq{eq:fitghost} and the other with the alternative ansatz
\beq
f_J^{(2)}(p) = b_1 +  \frac{b_2 \hat{p}^2}{(1+ \hat{p}^2)^{(1-\kappa)}},
\qquad \hat{p} \equiv p/m_{gh}.
\label{eq:fitghost2}
\eeq 
This function takes a nonzero value at $p=0$. We obtained good $\chi^2_{df}$ 
in both cases, 0.07 and 0.04, respectively. The parameters for both fitting 
curves are provided in the last two lines of \Tab{tab:fit_ghost}.

\subsection{Running coupling}
Taking the gluon dressing function results from \cite{Bornyakov:2009ug}
into account we can compute the minimal MOM scheme running coupling 
\cite{vonSmekal:2009ae} via
\beq
\alpha_s(p) = \frac{g_0^2}{4\pi} Z(p) J(p)^2 \,,
\label{eq:alpha_s}
\eeq
where $Z(p)$ and $J(p)$ are the bare gluon 
and ghost dressing functions, respectively.

\begin{figure*}[tb]
\centering
\includegraphics[width=5.9cm,angle=270]{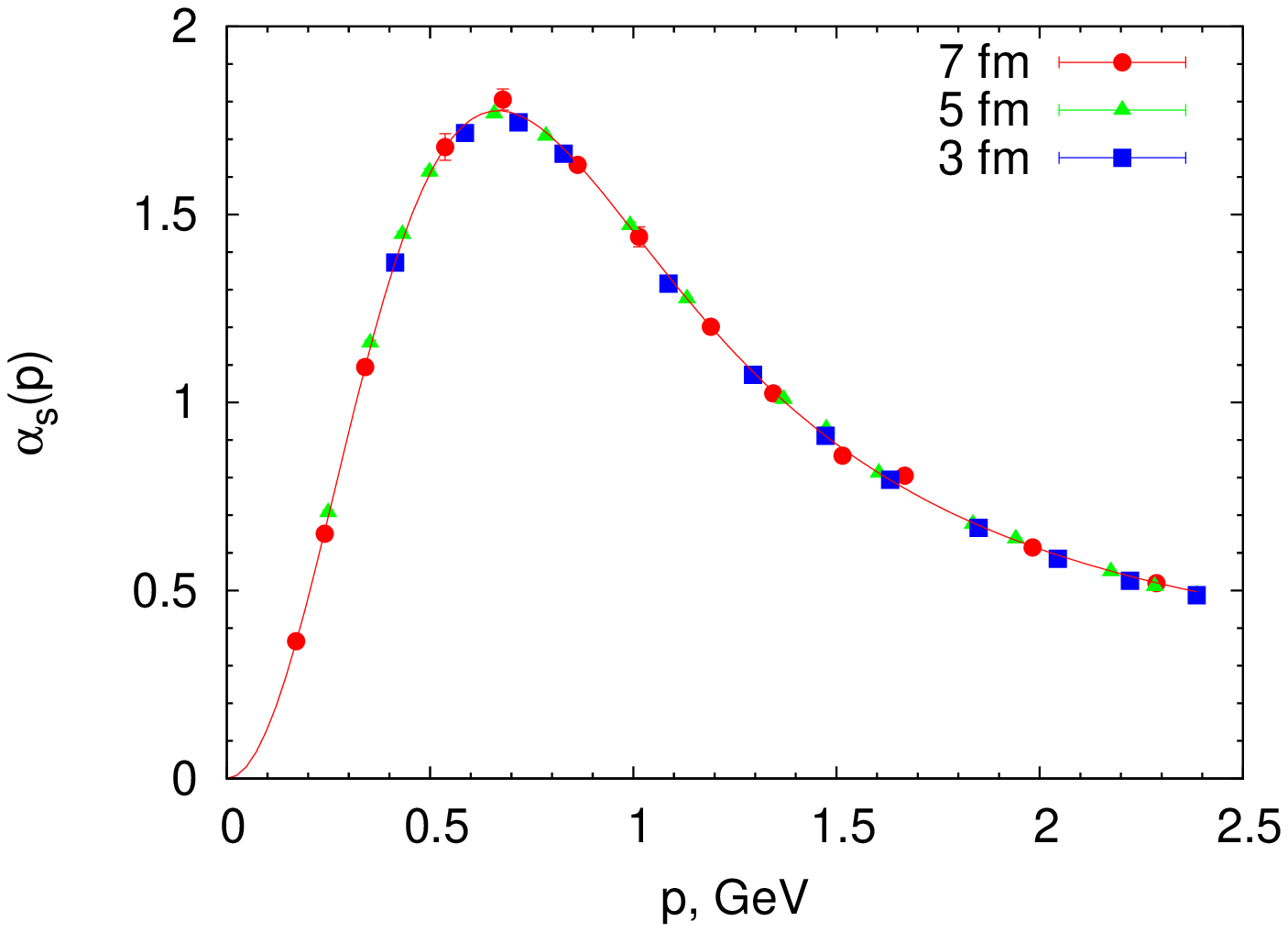}
\includegraphics[width=5.9cm,angle=270]{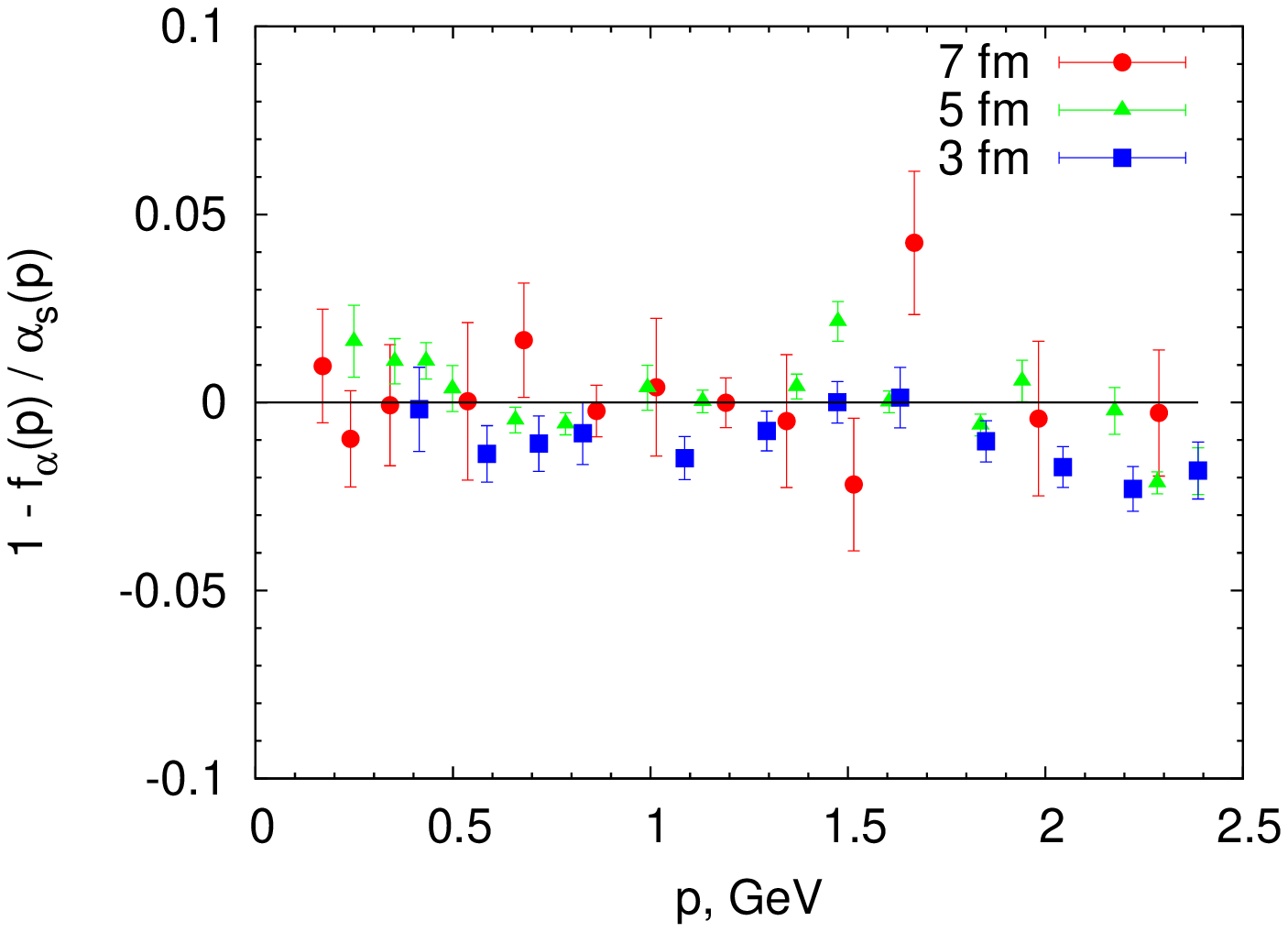}
\caption{{\bf Left:} The momentum dependence of the running coupling 
for three different lattice sizes at $\beta=2.3$. The curve shows a fit with
\Eq{eq:fitcoupl} for $aL \simeq 7$~fm. 
{\bf Right:} The relative deviation of the data 
for the running coupling $\alpha_s(p)$ from the fitting curve.
}
\label{fig:coupl_2p3}
\end{figure*}

\begin{figure*}[tb]
\centering
\includegraphics[width=5.9cm,angle=270]{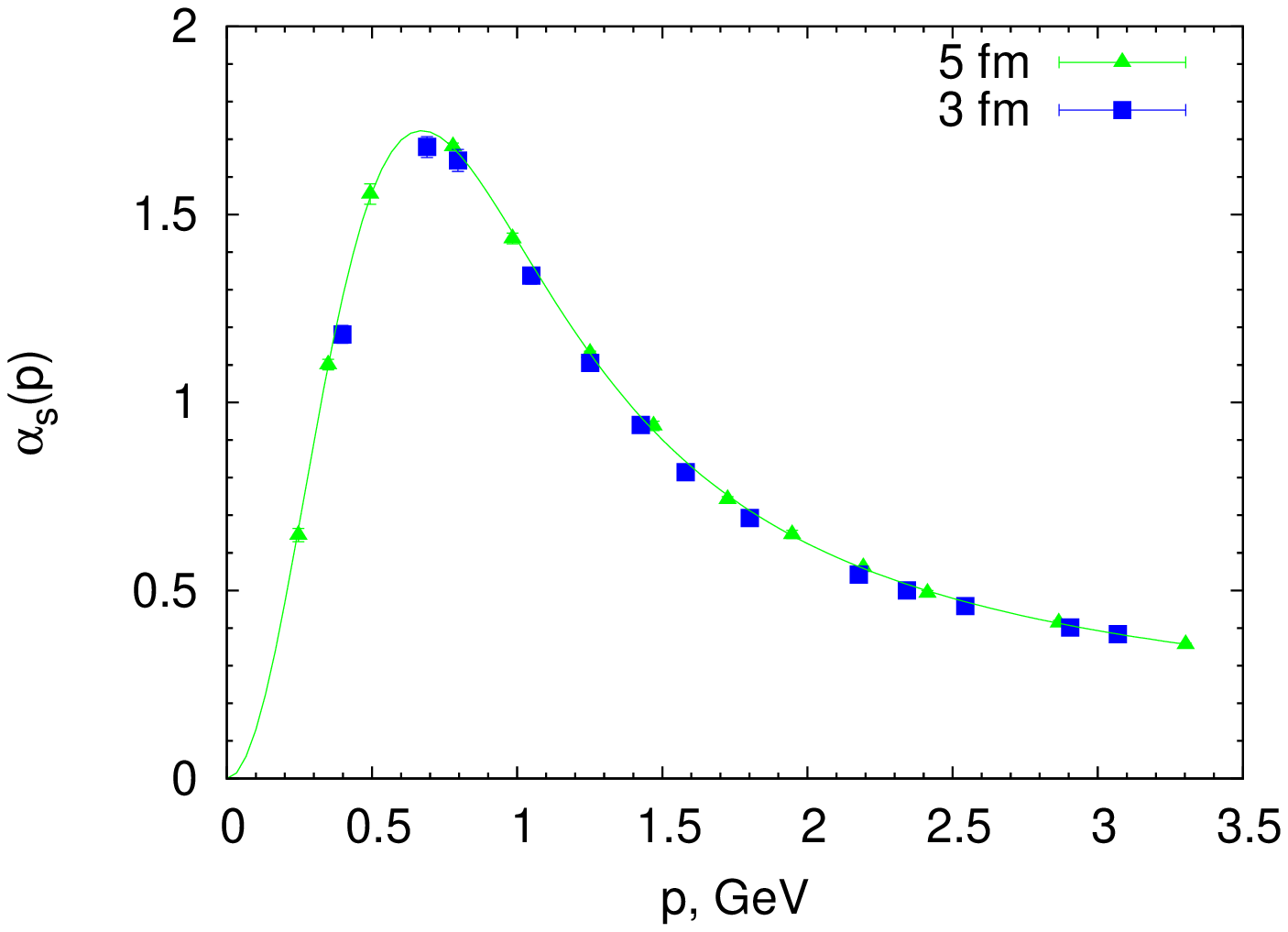}
\includegraphics[width=5.9cm,angle=270]{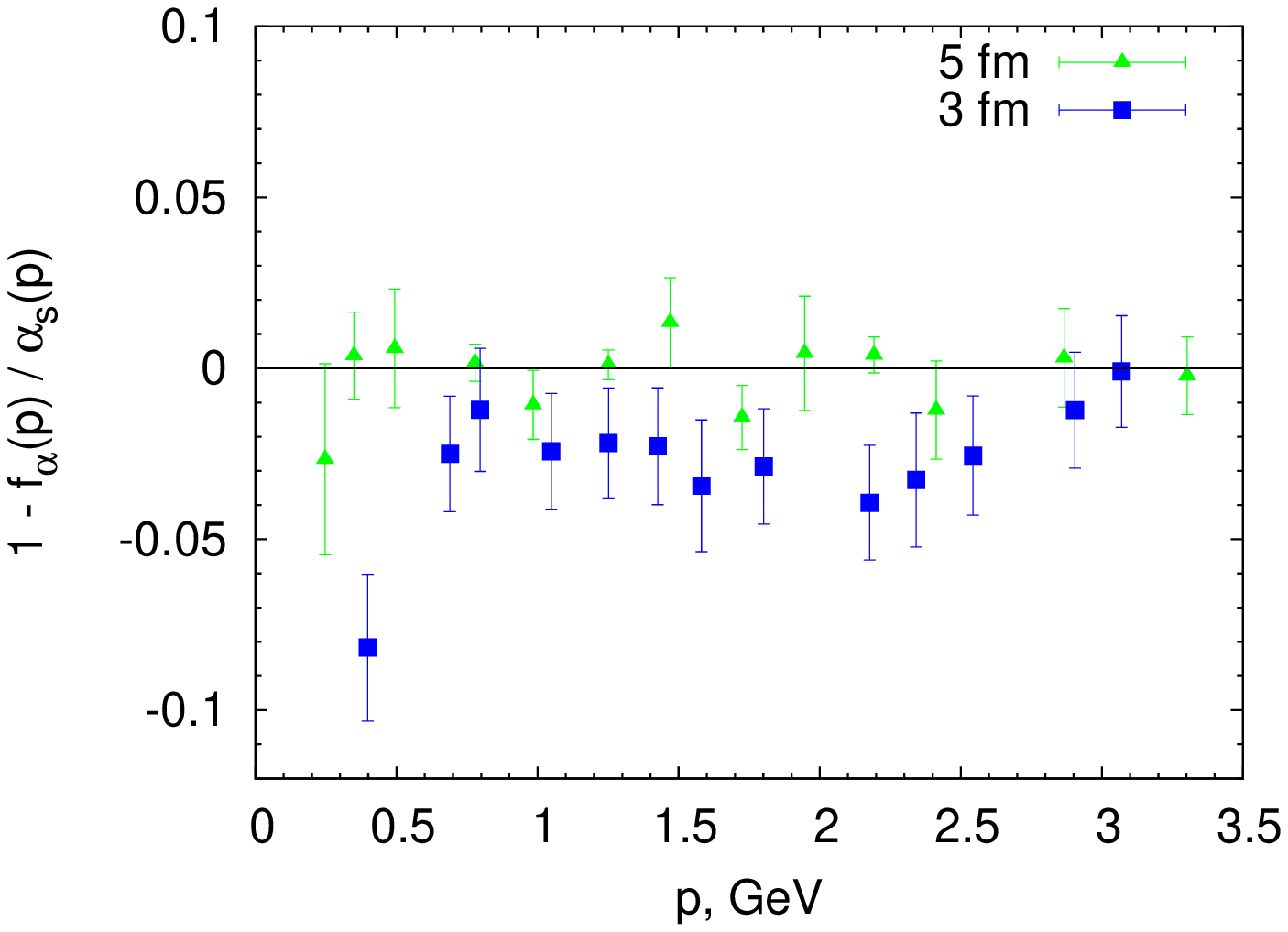}
\caption{Same as in \Fig{fig:coupl_2p3} but for two lattice sizes
at $\beta=2.4$. The curve shows the fit result with \Eq{eq:fitcoupl} 
for $aL \simeq 5$~fm.
}
\label{fig:coupl_2p4}
\end{figure*}

For the running coupling we use the following dimensionless fitting function:
\beq
f_\alpha(p) = \frac{c_1 \hat{p}^2}{1 + \hat{p}^2} + \frac{c_2 \hat{p}^2}{(1 + 
              \hat{p}^2)^2} + \frac{c_3 \hat{p}^2}{(1 + \hat{p}^2 )^4}\,,
\quad \hat{p} \equiv p / m_{\alpha}\,.
\label{eq:fitcoupl}
\eeq
The fit results for the same combinations of values $(\beta,~L)$ as for the 
ghost dressing function (see \Tab{tab:fit_ghost}) are collected in
\Tab{tab:fit_coupling}.

Finite volume effects for the running coupling are shown in 
\Fig{fig:coupl_2p3} for $\beta=2.3$ and in \Fig{fig:coupl_2p4} for $\beta=2.4$. 
In both cases one can see the finite volume effects to be reasonably small 
(less than 5\%) at a linear physical lattice extension $aL \simeq 3$~fm and 
for momenta $|p|\, \ageq \, 0.6$ GeV.

\begin{figure*}[tb]
\centering
\includegraphics[width=5.9cm,angle=270]{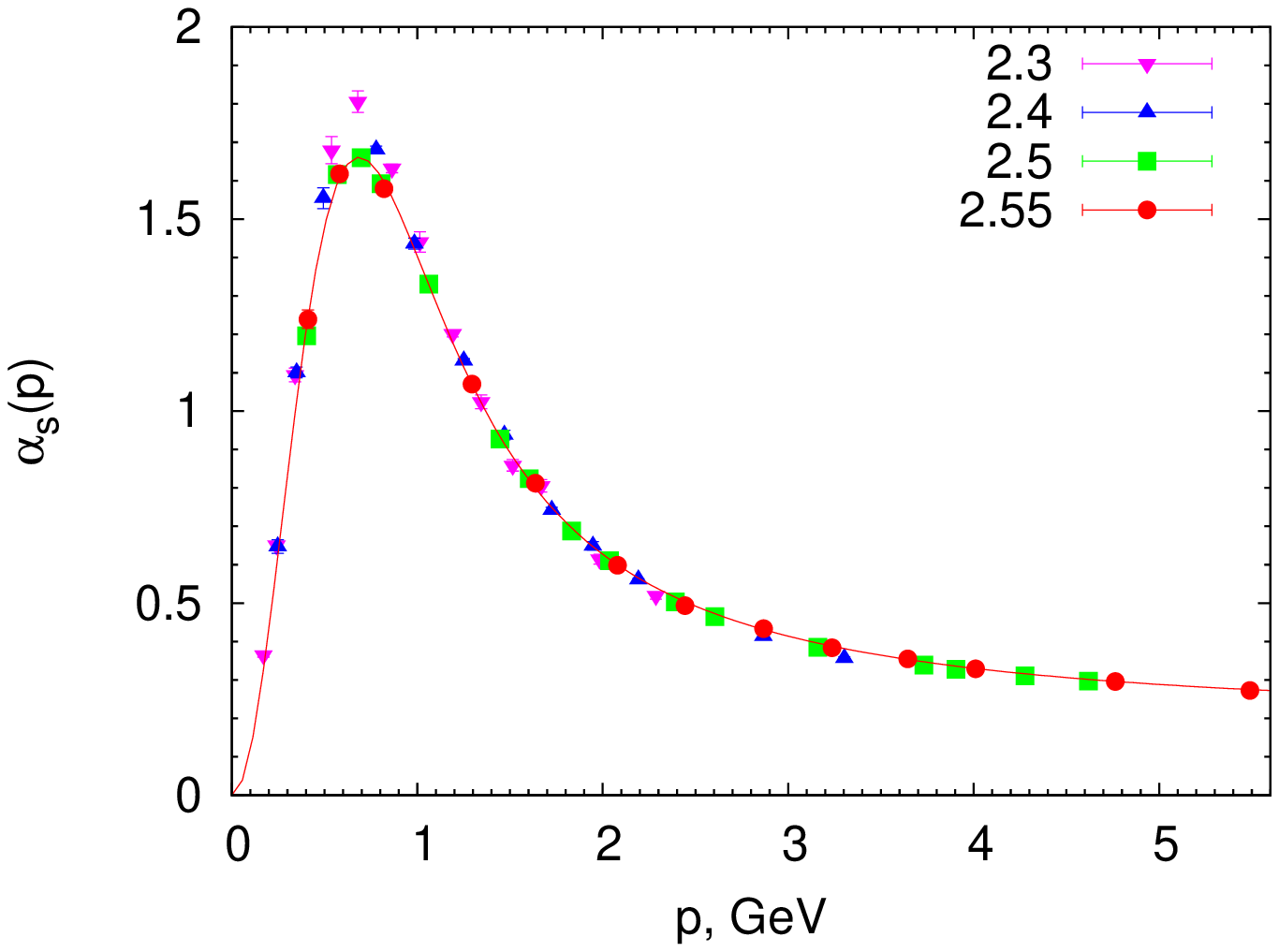}
\includegraphics[width=5.9cm,angle=270]{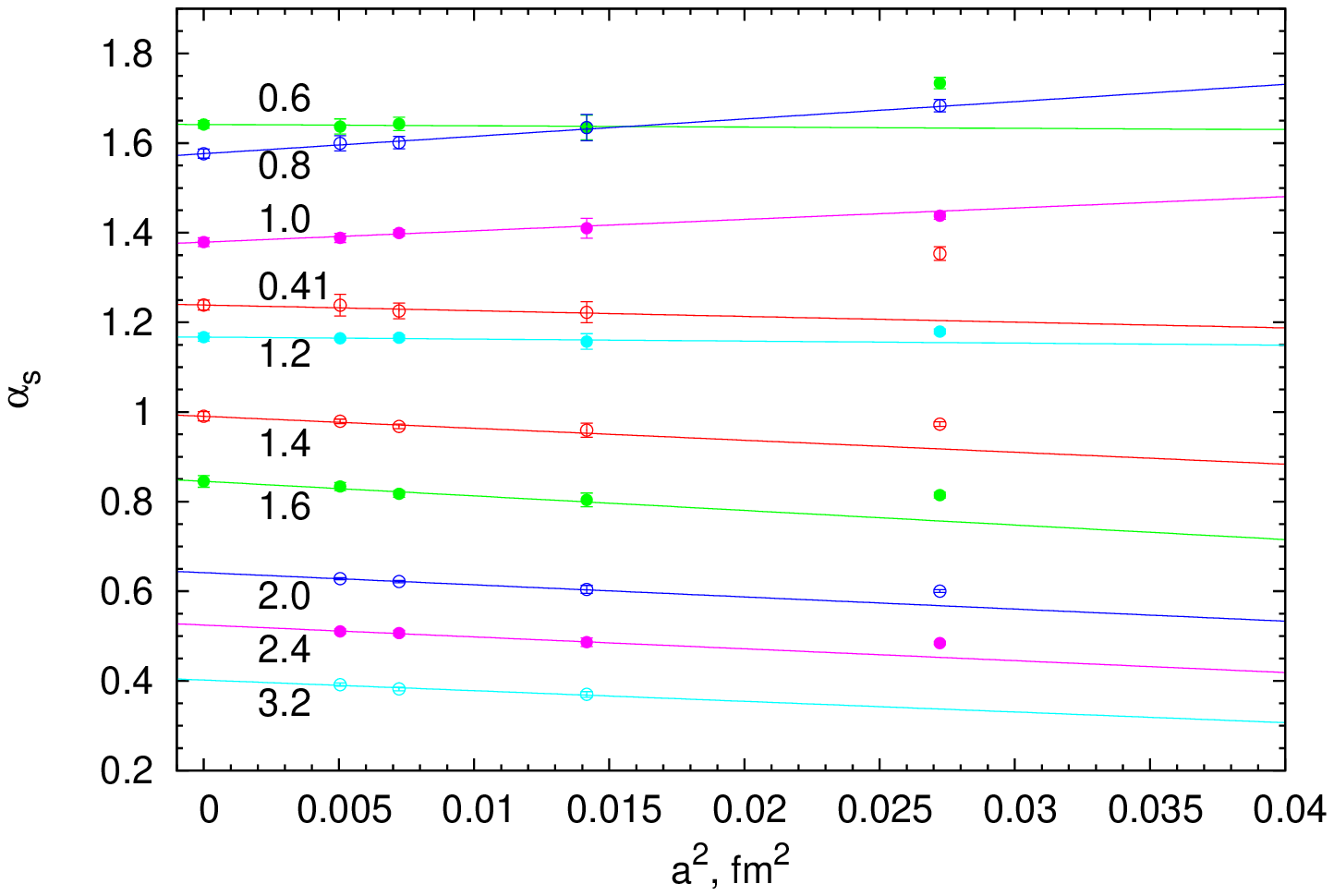}
\caption{{\bf Left:} The momentum dependence of the running coupling 
for four different $\beta$-values or lattice spacings at lattice 
extent $aL \simeq 3$~fm. The curve shows the fit result with 
\Eq{eq:fitcoupl} for $\beta=2.55$. {\bf Right:} The extrapolation to the 
continuum limit with a function linear in $a^2$ for selected momenta
(in GeV).
}
\label{fig:coupl_scal}
\end{figure*}

Results for the scaling check of $\alpha_s(p)$ taking into account four 
lattice spacings for the same extent of $aL \simeq 3$~fm are presented in 
\Fig{fig:coupl_scal}. We see relative deviations for $\beta=2.3$
in comparison with $\beta=2.55$ up to a $10\%$-level within the momentum 
range explored.
Similar to the ghost dressing function we made extrapolations to the 
continuum limit for selected momenta. The running coupling for these selected 
momenta as a linear function of $a^2$ is shown in the right panel of 
\Fig{fig:coupl_scal} together with extrapolations to the $a=0$ limit.
One can see that finite lattice spacing effects are very strong at $\beta=2.3$.
The respective data were not included into the continuum extrapolation.
Another feature seen from this figure is that the sign of the scaling violation
effects changes twice: it is negative to the left from the maximum of 
$\alpha_s(p)$ ($p=0.41$ GeV and $0.6$ GeV), becomes positive right above it 
($p=0.8$ MeV and $1.0$ MeV) and then again turns negative. In the range of momenta
$p > 1.2$ GeV the effect is rather stable in strength up to our maximal
momentum value.

\begin{figure}[tb]
\centering
\includegraphics[width=5.7cm,angle=270]{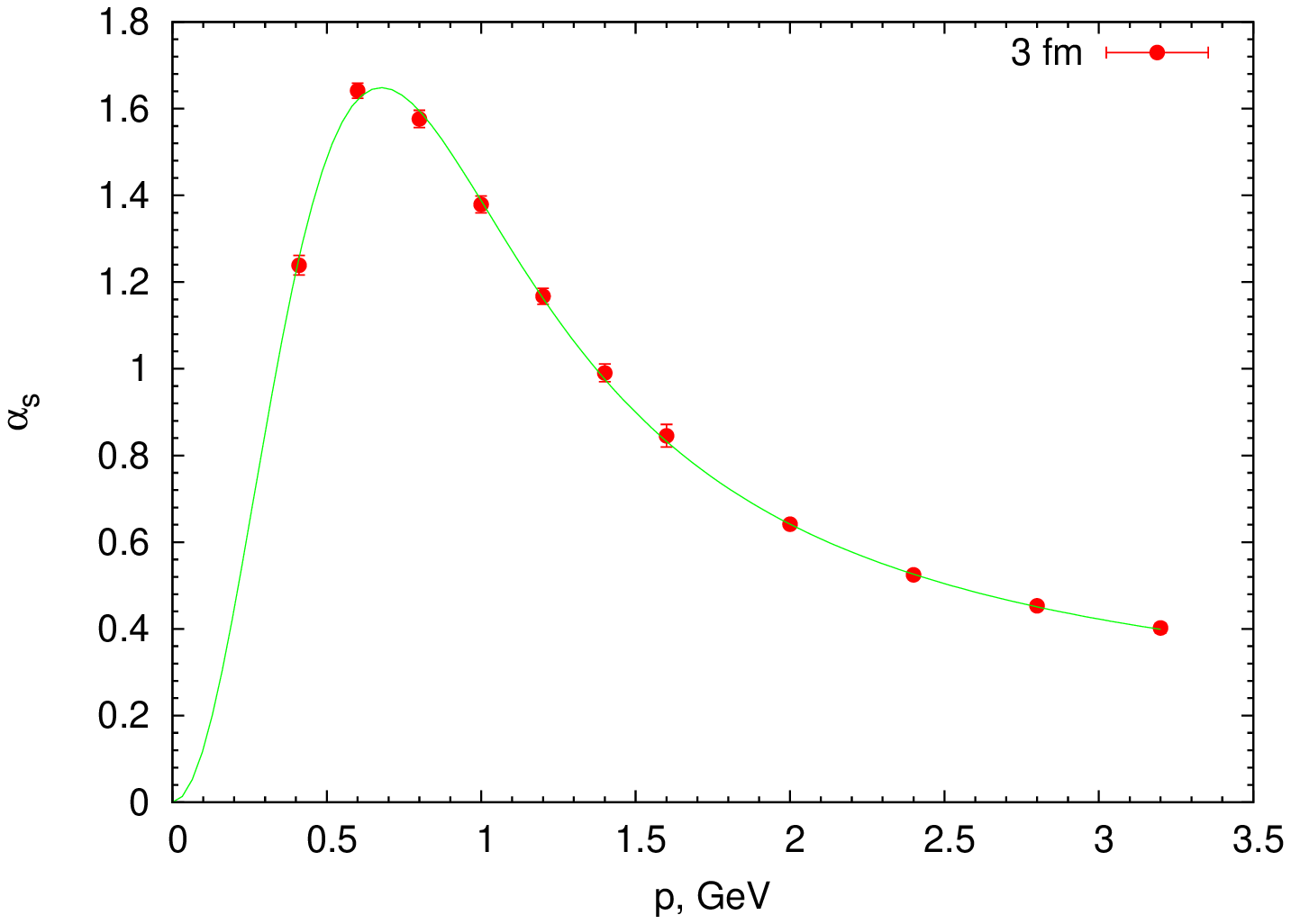}
\caption{The momentum dependence of the running coupling $\alpha_s(p)$
extracted in the continuum limit for selected momenta and $aL \simeq 3$~fm.  
The curve shows a fit corresponding to the ansatz \Eq{eq:fitcoupl}.
}
\label{fig:alphacont}
\end{figure}

In \Fig{fig:alphacont} we present our results for the continuum values
of the running coupling $\alpha_s(p)$ for linear size $aL \simeq 3$ fm.
For the fit the same ansatz according to \Eq{eq:fitcoupl} was used. 
The fit parameters are included in \Tab{tab:fit_coupling} (as the last line).

Since the running coupling $\alpha_s$ seems to tend to zero in the
IR limit, our results obtained within the framework of Landau
gauge fixing as described above are fully compatible with the IR-decoupling 
scenario discussed in the context of the Dyson-Schwinger and functional 
renormalization group approach \cite{Fischer:2008uz,Boucaud:2011ug}.  

\begin{table}[h]
\begin{center}
\vspace*{0.2cm}
\begin{tabular}{|c|c|c|c|c|c|c|} \hline
$\beta$ & $L$ & $m_{\alpha}$~[GeV] & $c_1$ & $c_2$ & $c_3$ & $\chi^2_{df}$ \\ 
\hline
 2.30  & 44 & 1.03(1) & 0.19(4)   & 2.3(2)  & 12(2)   & 0.95 \\
\hline 
 2.40  & 42 & 1.01(1) & 0.16(1)   & 2.63(5) & 11.0(7) & 0.81 \\
\hline
 2.55  & 42 & 1.04(1) & 0.205(3)  & 2.29(3) & 11.0(7) & 1.3  \\
\hline\hline
 c.l.  &    & 1.01(2) & 0.199(15) & 2.56(8) & 10.0(1.1) & 0.57 \\
\hline
\end{tabular}
\end{center}
\caption{Values of the fit parameters for $\alpha_s$ (\Eq{eq:fitcoupl}) 
and the corresponding $\chi^2_{df}$. 
The last line collects the corresponding fit parameter values in 
the continuum limit extra\-polated case for a linear lattice size of
$3$ fm.
} 
\label{tab:fit_coupling}
\end{table}

\section{Conclusions}
\label{sec:conclusions}

Completing an earlier work \cite{Bornyakov:2009ug} we have computed the 
Landau gauge ghost dressing function for lattice $SU(2)$ pure gauge theory. 
In combination with the former results for the gluon propagator 
we have now presented the running coupling in the minimal MOM scheme.
We have employed the same sets of gauge-fixed field configurations as 
analysed in \cite{Bornyakov:2009ug}.
They had been obtained with a gauge fixing method consisting of a combined 
application of $Z(2)$-flips and repeated simulated annealing with subsequent 
overrelaxation for the gauge functional. This method was used in order to get 
as close as possible to the fundamental modular region i.e. to the global 
extremum of the gauge functional, by choosing among ${\cal O}(50 - 80)$ copies.
It previously was suggested to provide a possible solution for the Gribov problem
with suppressed finite size effects \cite{Bogolubsky:2007bw,Bogolubsky:2007pq}. 

Assuming that the Gribov problem is kept under control to the best of our 
present knowledge, we concentrated
ourselves on systematic effects like finite size and lattice spacing
dependences. While finite size effects were confirmed to be rather small,
the lattice spacing artifacts turned out to be non-negligible
as well for the renormalized ghost dressing function as for the running
coupling. In both cases for a linear lattice size of approximately 
$aL \simeq 3$~fm (and for the ghost dressing function with a subtraction 
momentum of $\mu = 2.2$~GeV) we have seen relative deviations at $\beta=2.30$ 
from the results obtained at our largest $\beta=2.55$ reaching a 
ten percent level at the lowest accessible momentum values and still
around five percent in the non-perturbative region around 1~GeV.   
This tells us that lattice results for Landau gauge gluon and ghost 
propagators in this momentum range have still to be taken with some 
caution what concerns the Gribov problem and the continuum limit. 

Consequently we tried an extrapolation to the continuum limit for the 
$aL \simeq 3$ fm volume, where we could rely on several values of the 
lattice spacing.
We did this with fixed physical momenta chosen between $0.41$ GeV and 
$3.2$ GeV. We presented fit formulae for the continuum limit 
of the ghost dressing function as well as of the running coupling valid 
in this range. For momenta above $0.6$ GeV we have seen that also finite 
volume effects are under control.

Although the ghost dressing function in the restricted momentum range 
has been equally well fitted by a weakly IR singular behavior 
(see \Eq{eq:fitghost}) or with an IR regular ansatz \Eq{eq:fitghost2},
thus leaving open an IR finite limit, the result for the running coupling 
$\alpha_s(p)$ turned out to be robust, what 
emphasizes the compatibility with the infrared {\it decoupling solution} 
of Dyson-Schwinger and functional renormalization group equations.   

\vspace*{0.8cm}
\subsection*{Acknowledgments}
This investigation has been partly supported by the Heisenberg-Landau
program of collaboration between the Bogoliubov Laboratory of Theoretical 
Physics of the Joint Institute for Nuclear Research Dubna (Russia) and 
German institutes and partly by the DFG grant Mu 932/7-1.
V.G.B. acknowledges support by RFBR grant 11-02-01227-a 
and jointly with V.K.M. by RFBR grant 13-02-01387-a.

\bibliographystyle{apsrev}

\begin{thebibliography}{79}
\expandafter\ifx\csname natexlab\endcsname\relax\def\natexlab#1{#1}\fi
\expandafter\ifx\csname bibnamefont\endcsname\relax
  \def\bibnamefont#1{#1}\fi
\expandafter\ifx\csname bibfnamefont\endcsname\relax
  \def\bibfnamefont#1{#1}\fi
\expandafter\ifx\csname citenamefont\endcsname\relax
  \def\citenamefont#1{#1}\fi
\expandafter\ifx\csname url\endcsname\relax
  \def\url#1{\texttt{#1}}\fi
\expandafter\ifx\csname urlprefix\endcsname\relax\def\urlprefix{URL }\fi
\providecommand{\bibinfo}[2]{#2}
\providecommand{\eprint}[2][]{\url{#2}}

\bibitem[{\citenamefont{Gribov}(1978)}]{Gribov:1977wm}
\bibinfo{author}{\bibfnamefont{V.~N.} \bibnamefont{Gribov}},
  \bibinfo{journal}{Nucl. Phys.} \textbf{\bibinfo{volume}{B139}},
  \bibinfo{pages}{1} (\bibinfo{year}{1978}).

\bibitem[{\citenamefont{Kugo and Ojima}(1979)}]{Kugo:1979gm}
\bibinfo{author}{\bibfnamefont{T.}~\bibnamefont{Kugo}} \bibnamefont{and}
  \bibinfo{author}{\bibfnamefont{I.}~\bibnamefont{Ojima}},
  \bibinfo{journal}{Prog. Theor. Phys. Suppl.} \textbf{\bibinfo{volume}{66}},
  \bibinfo{pages}{1} (\bibinfo{year}{1979}).

\bibitem[{\citenamefont{Zwanziger}(1991)}]{Zwanziger:1991gz}
\bibinfo{author}{\bibfnamefont{D.}~\bibnamefont{Zwanziger}},
  \bibinfo{journal}{Nucl. Phys.} \textbf{\bibinfo{volume}{B364}},
  \bibinfo{pages}{127} (\bibinfo{year}{1991}).

\bibitem[{\citenamefont{Zwanziger}(2002)}]{Zwanziger:2001kw}
\bibinfo{author}{\bibfnamefont{D.}~\bibnamefont{Zwanziger}},
  \bibinfo{journal}{Phys. Rev.} \textbf{\bibinfo{volume}{D65}},
  \bibinfo{pages}{094039} (\bibinfo{year}{2002}), \eprint{hep-th/0109224}.

\bibitem[{\citenamefont{Zwanziger}(2009)}]{Zwanziger:2009je}
\bibinfo{author}{\bibfnamefont{D.}~\bibnamefont{Zwanziger}}
  (\bibinfo{year}{2009}), \eprint{0904.2380}.

\bibitem[{\citenamefont{von Smekal et~al.}(1997)\citenamefont{von Smekal,
  Alkofer, and Hauck}}]{vonSmekal:1997is}
\bibinfo{author}{\bibfnamefont{L.}~\bibnamefont{von Smekal}},
  \bibinfo{author}{\bibfnamefont{R.}~\bibnamefont{Alkofer}}, \bibnamefont{and}
  \bibinfo{author}{\bibfnamefont{A.}~\bibnamefont{Hauck}},
  \bibinfo{journal}{Phys. Rev. Lett.} \textbf{\bibinfo{volume}{79}},
  \bibinfo{pages}{3591} (\bibinfo{year}{1997}), \eprint{hep-ph/9705242}.

\bibitem[{\citenamefont{von Smekal et~al.}(1998)\citenamefont{von Smekal,
  Hauck, and Alkofer}}]{vonSmekal:1997vx}
\bibinfo{author}{\bibfnamefont{L.}~\bibnamefont{von Smekal}},
  \bibinfo{author}{\bibfnamefont{A.}~\bibnamefont{Hauck}}, \bibnamefont{and}
  \bibinfo{author}{\bibfnamefont{R.}~\bibnamefont{Alkofer}},
  \bibinfo{journal}{Ann. Phys.} \textbf{\bibinfo{volume}{267}},
  \bibinfo{pages}{1} (\bibinfo{year}{1998}), \eprint{hep-ph/9707327}.

\bibitem[{\citenamefont{Alkofer and von Smekal}(2001)}]{Alkofer:2000wg}
\bibinfo{author}{\bibfnamefont{R.}~\bibnamefont{Alkofer}} \bibnamefont{and}
  \bibinfo{author}{\bibfnamefont{L.}~\bibnamefont{von Smekal}},
  \bibinfo{journal}{Phys. Rept.} \textbf{\bibinfo{volume}{353}},
  \bibinfo{pages}{281} (\bibinfo{year}{2001}), \eprint{hep-ph/0007355}.

\bibitem[{\citenamefont{Alkofer et~al.}(2003)\citenamefont{Alkofer, Fischer,
  and von Smekal}}]{Alkofer:2002aa}
\bibinfo{author}{\bibfnamefont{R.}~\bibnamefont{Alkofer}},
  \bibinfo{author}{\bibfnamefont{C.~S.} \bibnamefont{Fischer}},
  \bibnamefont{and} \bibinfo{author}{\bibfnamefont{L.}~\bibnamefont{von
  Smekal}}, \bibinfo{journal}{Eur. Phys. J.} \textbf{\bibinfo{volume}{A17}},
  \bibinfo{pages}{773} (\bibinfo{year}{2003}), \eprint{hep-ph/0209366}.

\bibitem[{\citenamefont{Fischer and Alkofer}(2002)}]{Fischer:2002hna}
\bibinfo{author}{\bibfnamefont{C.~S.} \bibnamefont{Fischer}} \bibnamefont{and}
  \bibinfo{author}{\bibfnamefont{R.}~\bibnamefont{Alkofer}},
  \bibinfo{journal}{Phys. Lett.} \textbf{\bibinfo{volume}{B536}},
  \bibinfo{pages}{177} (\bibinfo{year}{2002}), \eprint{hep-ph/0202202}.

\bibitem[{\citenamefont{Fischer et~al.}(2002)\citenamefont{Fischer, Alkofer,
  and Reinhardt}}]{Fischer:2002eq}
\bibinfo{author}{\bibfnamefont{C.~S.} \bibnamefont{Fischer}},
  \bibinfo{author}{\bibfnamefont{R.}~\bibnamefont{Alkofer}}, \bibnamefont{and}
  \bibinfo{author}{\bibfnamefont{H.}~\bibnamefont{Reinhardt}},
  \bibinfo{journal}{Phys. Rev.} \textbf{\bibinfo{volume}{D65}},
  \bibinfo{pages}{094008} (\bibinfo{year}{2002}), \eprint{hep-ph/0202195}.

\bibitem[{\citenamefont{Pawlowski et~al.}(2004)\citenamefont{Pawlowski, Litim,
  Nedelko, and von Smekal}}]{Pawlowski:2003hq}
\bibinfo{author}{\bibfnamefont{J.~M.} \bibnamefont{Pawlowski}},
  \bibinfo{author}{\bibfnamefont{D.~F.} \bibnamefont{Litim}},
  \bibinfo{author}{\bibfnamefont{S.}~\bibnamefont{Nedelko}}, \bibnamefont{and}
  \bibinfo{author}{\bibfnamefont{L.}~\bibnamefont{von Smekal}},
  \bibinfo{journal}{Phys. Rev. Lett.} \textbf{\bibinfo{volume}{93}},
  \bibinfo{pages}{152002} (\bibinfo{year}{2004}), \eprint{hep-th/0312324}.

\bibitem[{\citenamefont{Alkofer et~al.}(2005)\citenamefont{Alkofer, Fischer,
  and Llanes-Estrada}}]{Alkofer:2004it}
\bibinfo{author}{\bibfnamefont{R.}~\bibnamefont{Alkofer}},
  \bibinfo{author}{\bibfnamefont{C.~S.} \bibnamefont{Fischer}},
  \bibnamefont{and} \bibinfo{author}{\bibfnamefont{F.~J.}
  \bibnamefont{Llanes-Estrada}}, \bibinfo{journal}{Phys. Lett.}
  \textbf{\bibinfo{volume}{B611}}, \bibinfo{pages}{279} (\bibinfo{year}{2005}),
  \eprint{hep-th/0412330}.

\bibitem[{\citenamefont{Fischer and Pawlowski}(2007)}]{Fischer:2006vf}
\bibinfo{author}{\bibfnamefont{C.~S.} \bibnamefont{Fischer}} \bibnamefont{and}
  \bibinfo{author}{\bibfnamefont{J.~M.} \bibnamefont{Pawlowski}},
  \bibinfo{journal}{Phys. Rev.} \textbf{\bibinfo{volume}{D75}},
  \bibinfo{pages}{025012} (\bibinfo{year}{2007}), \eprint{hep-th/0609009}.

\bibitem[{\citenamefont{Boucaud et~al.}(2007)\citenamefont{Boucaud, Leroy,
  Le~Yaouanc, Lokhov, Micheli, Pene, Rodriguez-Quintero, and
  Roiesnel}}]{Boucaud:2007va}
\bibinfo{author}{\bibfnamefont{P.}~\bibnamefont{Boucaud}},
  \bibinfo{author}{\bibfnamefont{J.~P.} \bibnamefont{Leroy}},
  \bibinfo{author}{\bibfnamefont{A.}~\bibnamefont{Le~Yaouanc}},
  \bibinfo{author}{\bibfnamefont{A.~Y.} \bibnamefont{Lokhov}},
  \bibinfo{author}{\bibfnamefont{J.}~\bibnamefont{Micheli}},
  \bibinfo{author}{\bibfnamefont{O.}~\bibnamefont{Pene}},
  \bibinfo{author}{\bibfnamefont{J.}~\bibnamefont{Rodriguez-Quintero}},
  \bibnamefont{and} \bibinfo{author}{\bibfnamefont{C.}~\bibnamefont{Roiesnel}},
  \bibinfo{journal}{Eur. Phys. J.} \textbf{\bibinfo{volume}{A31}},
  \bibinfo{pages}{750} (\bibinfo{year}{2007}), \eprint{hep-ph/0701114}.

\bibitem[{\citenamefont{Boucaud
  et~al.}(2008{\natexlab{a}})\citenamefont{Boucaud, Leroy, Yaouanc, Micheli,
  Pene, and Rodriguez-Quintero}}]{Boucaud:2008ji}
\bibinfo{author}{\bibfnamefont{P.}~\bibnamefont{Boucaud}},
  \bibinfo{author}{\bibfnamefont{J.-P.} \bibnamefont{Leroy}},
  \bibinfo{author}{\bibfnamefont{A.~L.} \bibnamefont{Yaouanc}},
  \bibinfo{author}{\bibfnamefont{J.}~\bibnamefont{Micheli}},
  \bibinfo{author}{\bibfnamefont{O.}~\bibnamefont{Pene}}, \bibnamefont{and}
  \bibinfo{author}{\bibfnamefont{J.}~\bibnamefont{Rodriguez-Quintero}},
  \bibinfo{journal}{JHEP} \textbf{\bibinfo{volume}{06}}, \bibinfo{pages}{012}
  (\bibinfo{year}{2008}{\natexlab{a}}), \eprint{0801.2721}.

\bibitem[{\citenamefont{Boucaud
  et~al.}(2008{\natexlab{b}})\citenamefont{Boucaud, Leroy, Le~Yaouanc, Micheli,
  Pene, and Rodriguez-Quintero}}]{Boucaud:2008ky}
\bibinfo{author}{\bibfnamefont{P.}~\bibnamefont{Boucaud}},
  \bibinfo{author}{\bibfnamefont{J.~P.} \bibnamefont{Leroy}},
  \bibinfo{author}{\bibfnamefont{A.}~\bibnamefont{Le~Yaouanc}},
  \bibinfo{author}{\bibfnamefont{J.}~\bibnamefont{Micheli}},
  \bibinfo{author}{\bibfnamefont{O.}~\bibnamefont{Pene}}, \bibnamefont{and}
  \bibinfo{author}{\bibfnamefont{J.}~\bibnamefont{Rodriguez-Quintero}},
  \bibinfo{journal}{JHEP} \textbf{\bibinfo{volume}{06}}, \bibinfo{pages}{099}
  (\bibinfo{year}{2008}{\natexlab{b}}), \eprint{0803.2161}.

\bibitem[{\citenamefont{Aguilar et~al.}(2008)\citenamefont{Aguilar, Binosi, and
  Papavassiliou}}]{Aguilar:2008xm}
\bibinfo{author}{\bibfnamefont{A.~C.} \bibnamefont{Aguilar}},
  \bibinfo{author}{\bibfnamefont{D.}~\bibnamefont{Binosi}}, \bibnamefont{and}
  \bibinfo{author}{\bibfnamefont{J.}~\bibnamefont{Papavassiliou}},
  \bibinfo{journal}{Phys. Rev.} \textbf{\bibinfo{volume}{D78}},
  \bibinfo{pages}{025010} (\bibinfo{year}{2008}), \eprint{0802.1870}.

\bibitem[{\citenamefont{Pennington and Wilson}(2011)}]{Pennington:2011xs}
\bibinfo{author}{\bibfnamefont{M.}~\bibnamefont{Pennington}} \bibnamefont{and}
  \bibinfo{author}{\bibfnamefont{D.}~\bibnamefont{Wilson}},
  \bibinfo{journal}{Phys.Rev.} \textbf{\bibinfo{volume}{D84}},
  \bibinfo{pages}{119901} (\bibinfo{year}{2011}), \eprint{1109.2117}.

\bibitem[{\citenamefont{Fischer et~al.}(2009)\citenamefont{Fischer, Maas, and
  Pawlowski}}]{Fischer:2008uz}
\bibinfo{author}{\bibfnamefont{C.~S.} \bibnamefont{Fischer}},
  \bibinfo{author}{\bibfnamefont{A.}~\bibnamefont{Maas}}, \bibnamefont{and}
  \bibinfo{author}{\bibfnamefont{J.~M.} \bibnamefont{Pawlowski}},
  \bibinfo{journal}{Annals Phys.} \textbf{\bibinfo{volume}{324}},
  \bibinfo{pages}{2408} (\bibinfo{year}{2009}), \eprint{0810.1987}.

\bibitem[{\citenamefont{Braun et~al.}(2010)\citenamefont{Braun, Gies, and
  Pawlowski}}]{Braun:2007bx}
\bibinfo{author}{\bibfnamefont{J.}~\bibnamefont{Braun}},
  \bibinfo{author}{\bibfnamefont{H.}~\bibnamefont{Gies}}, \bibnamefont{and}
  \bibinfo{author}{\bibfnamefont{J.~M.} \bibnamefont{Pawlowski}},
  \bibinfo{journal}{Phys. Lett.} \textbf{\bibinfo{volume}{B684}},
  \bibinfo{pages}{262} (\bibinfo{year}{2010}), \eprint{0708.2413}.

\bibitem[{\citenamefont{Dudal et~al.}(2008{\natexlab{a}})\citenamefont{Dudal,
  Sorella, Vandersickel, and Verschelde}}]{Dudal:2007cw}
\bibinfo{author}{\bibfnamefont{D.}~\bibnamefont{Dudal}},
  \bibinfo{author}{\bibfnamefont{S.~P.} \bibnamefont{Sorella}},
  \bibinfo{author}{\bibfnamefont{N.}~\bibnamefont{Vandersickel}},
  \bibnamefont{and}
  \bibinfo{author}{\bibfnamefont{H.}~\bibnamefont{Verschelde}},
  \bibinfo{journal}{Phys. Rev.} \textbf{\bibinfo{volume}{D77}},
  \bibinfo{pages}{071501} (\bibinfo{year}{2008}{\natexlab{a}}),
  \eprint{0711.4496}.

\bibitem[{\citenamefont{Dudal et~al.}(2008{\natexlab{b}})\citenamefont{Dudal,
  Gracey, Sorella, Vandersickel, and Verschelde}}]{Dudal:2008sp}
\bibinfo{author}{\bibfnamefont{D.}~\bibnamefont{Dudal}},
  \bibinfo{author}{\bibfnamefont{J.~A.} \bibnamefont{Gracey}},
  \bibinfo{author}{\bibfnamefont{S.~P.} \bibnamefont{Sorella}},
  \bibinfo{author}{\bibfnamefont{N.}~\bibnamefont{Vandersickel}},
  \bibnamefont{and}
  \bibinfo{author}{\bibfnamefont{H.}~\bibnamefont{Verschelde}},
  \bibinfo{journal}{Phys. Rev.} \textbf{\bibinfo{volume}{D78}},
  \bibinfo{pages}{065047} (\bibinfo{year}{2008}{\natexlab{b}}),
  \eprint{0806.4348}.

\bibitem[{\citenamefont{Alkofer}(2007)}]{Alkofer:2006jf}
\bibinfo{author}{\bibfnamefont{R.}~\bibnamefont{Alkofer}},
  \bibinfo{journal}{Braz. J. Phys.} \textbf{\bibinfo{volume}{37}},
  \bibinfo{pages}{144} (\bibinfo{year}{2007}), \eprint{hep-ph/0611090}.

\bibitem[{\citenamefont{Eichmann et~al.}(2009)\citenamefont{Eichmann, Cloet,
  Alkofer, Krassnigg, and Roberts}}]{Eichmann:2008ef}
\bibinfo{author}{\bibfnamefont{G.}~\bibnamefont{Eichmann}},
  \bibinfo{author}{\bibfnamefont{I.~C.} \bibnamefont{Cloet}},
  \bibinfo{author}{\bibfnamefont{R.}~\bibnamefont{Alkofer}},
  \bibinfo{author}{\bibfnamefont{A.}~\bibnamefont{Krassnigg}},
  \bibnamefont{and} \bibinfo{author}{\bibfnamefont{C.~D.}
  \bibnamefont{Roberts}}, \bibinfo{journal}{Phys. Rev.}
  \textbf{\bibinfo{volume}{C79}}, \bibinfo{pages}{012202}
  (\bibinfo{year}{2009}), \eprint{0810.1222}.

\bibitem[{\citenamefont{Sternbeck et~al.}(2012)\citenamefont{Sternbeck,
  Maltman, M{\"u}ller-Preussker, and von Smekal}}]{Sternbeck:2012qs}
\bibinfo{author}{\bibfnamefont{A.}~\bibnamefont{Sternbeck}},
  \bibinfo{author}{\bibfnamefont{K.}~\bibnamefont{Maltman}},
  \bibinfo{author}{\bibfnamefont{M.}~\bibnamefont{M{\"u}ller-Preussker}},
  \bibnamefont{and} \bibinfo{author}{\bibfnamefont{L.}~\bibnamefont{von
  Smekal}}, \bibinfo{journal}{PoS} \textbf{\bibinfo{volume}{LATTICE2012}},
  \bibinfo{pages}{243} (\bibinfo{year}{2012}), \eprint{1212.2039}.

\bibitem[{\citenamefont{Burger et~al.}(2013)\citenamefont{Burger, Lubicz,
  M{\"u}ller-Preussker, Simula, and Urbach}}]{Burger:2012ti}
\bibinfo{author}{\bibfnamefont{F.}~\bibnamefont{Burger}},
  \bibinfo{author}{\bibfnamefont{V.}~\bibnamefont{Lubicz}},
  \bibinfo{author}{\bibfnamefont{M.}~\bibnamefont{M{\"u}ller-Preussker}},
  \bibinfo{author}{\bibfnamefont{S.}~\bibnamefont{Simula}}, \bibnamefont{and}
  \bibinfo{author}{\bibfnamefont{C.}~\bibnamefont{Urbach}},
  \bibinfo{journal}{Phys. Rev.} \textbf{\bibinfo{volume}{D87}},
  \bibinfo{pages}{034514} (\bibinfo{year}{2013}), \eprint{1210.0838}.

\bibitem[{\citenamefont{Chetyrkin}(2005)}]{Chetyrkin:2004mf}
\bibinfo{author}{\bibfnamefont{K.~G.} \bibnamefont{Chetyrkin}},
  \bibinfo{journal}{Nucl. Phys.} \textbf{\bibinfo{volume}{B710}},
  \bibinfo{pages}{499} (\bibinfo{year}{2005}), \eprint{hep-ph/0405193}.

\bibitem[{\citenamefont{Chetyrkin and Maier}(2009)}]{Chetyrkin:2009kh}
\bibinfo{author}{\bibfnamefont{K.~G.} \bibnamefont{Chetyrkin}}
  \bibnamefont{and} \bibinfo{author}{\bibfnamefont{A.}~\bibnamefont{Maier}}
  (\bibinfo{year}{2009}), \eprint{0911.0594}.

\bibitem[{\citenamefont{Mandula and Ogilvie}(1987)}]{Mandula:1987rh}
\bibinfo{author}{\bibfnamefont{J.~E.} \bibnamefont{Mandula}} \bibnamefont{and}
  \bibinfo{author}{\bibfnamefont{M.}~\bibnamefont{Ogilvie}},
  \bibinfo{journal}{Phys. Lett.} \textbf{\bibinfo{volume}{B185}},
  \bibinfo{pages}{127} (\bibinfo{year}{1987}).

\bibitem[{\citenamefont{Suman and Schilling}(1996)}]{Suman:1995zg}
\bibinfo{author}{\bibfnamefont{H.}~\bibnamefont{Suman}} \bibnamefont{and}
  \bibinfo{author}{\bibfnamefont{K.}~\bibnamefont{Schilling}},
  \bibinfo{journal}{Phys. Lett.} \textbf{\bibinfo{volume}{B373}},
  \bibinfo{pages}{314} (\bibinfo{year}{1996}), \eprint{hep-lat/9512003}.

\bibitem[{\citenamefont{Leinweber et~al.}(1999)\citenamefont{Leinweber,
  Skullerud, Williams, and Parrinello}}]{Leinweber:1998uu}
\bibinfo{author}{\bibfnamefont{D.~B.} \bibnamefont{Leinweber}},
  \bibinfo{author}{\bibfnamefont{J.~I.} \bibnamefont{Skullerud}},
  \bibinfo{author}{\bibfnamefont{A.~G.} \bibnamefont{Williams}},
  \bibnamefont{and}
  \bibinfo{author}{\bibfnamefont{C.}~\bibnamefont{Parrinello}}
  (\bibinfo{collaboration}{UKQCD}), \bibinfo{journal}{Phys. Rev.}
  \textbf{\bibinfo{volume}{D60}}, \bibinfo{pages}{094507}
  (\bibinfo{year}{1999}), \eprint{hep-lat/9811027}.

\bibitem[{\citenamefont{Becirevic et~al.}(1999)\citenamefont{Becirevic,
  Boucaud, Leroy, Micheli, Pene, Rodriguez-Quintero, and
  Roiesnel}}]{Becirevic:1999uc}
\bibinfo{author}{\bibfnamefont{D.}~\bibnamefont{Becirevic}},
  \bibinfo{author}{\bibfnamefont{P.}~\bibnamefont{Boucaud}},
  \bibinfo{author}{\bibfnamefont{J.~P.} \bibnamefont{Leroy}},
  \bibinfo{author}{\bibfnamefont{J.}~\bibnamefont{Micheli}},
  \bibinfo{author}{\bibfnamefont{O.}~\bibnamefont{Pene}},
  \bibinfo{author}{\bibfnamefont{J.}~\bibnamefont{Rodriguez-Quintero}},
  \bibnamefont{and} \bibinfo{author}{\bibfnamefont{C.}~\bibnamefont{Roiesnel}},
  \bibinfo{journal}{Phys. Rev.} \textbf{\bibinfo{volume}{D60}},
  \bibinfo{pages}{094509} (\bibinfo{year}{1999}), \eprint{hep-ph/9903364}.

\bibitem[{\citenamefont{Becirevic et~al.}(2000)\citenamefont{Becirevic,
  Boucaud, Leroy, Micheli, Pene, Rodriguez-Quintero, and
  Roiesnel}}]{Becirevic:1999hj}
\bibinfo{author}{\bibfnamefont{D.}~\bibnamefont{Becirevic}},
  \bibinfo{author}{\bibfnamefont{P.}~\bibnamefont{Boucaud}},
  \bibinfo{author}{\bibfnamefont{J.~P.} \bibnamefont{Leroy}},
  \bibinfo{author}{\bibfnamefont{J.}~\bibnamefont{Micheli}},
  \bibinfo{author}{\bibfnamefont{O.}~\bibnamefont{Pene}},
  \bibinfo{author}{\bibfnamefont{J.}~\bibnamefont{Rodriguez-Quintero}},
  \bibnamefont{and} \bibinfo{author}{\bibfnamefont{C.}~\bibnamefont{Roiesnel}},
  \bibinfo{journal}{Phys. Rev.} \textbf{\bibinfo{volume}{D61}},
  \bibinfo{pages}{114508} (\bibinfo{year}{2000}), \eprint{hep-ph/9910204}.

\bibitem[{\citenamefont{Bonnet et~al.}(2000)\citenamefont{Bonnet, Bowman,
  Leinweber, and Williams}}]{Bonnet:2000kw}
\bibinfo{author}{\bibfnamefont{F.~D.~R.} \bibnamefont{Bonnet}},
  \bibinfo{author}{\bibfnamefont{P.~O.} \bibnamefont{Bowman}},
  \bibinfo{author}{\bibfnamefont{D.~B.} \bibnamefont{Leinweber}},
  \bibnamefont{and} \bibinfo{author}{\bibfnamefont{A.~G.}
  \bibnamefont{Williams}}, \bibinfo{journal}{Phys. Rev.}
  \textbf{\bibinfo{volume}{D62}}, \bibinfo{pages}{051501}
  (\bibinfo{year}{2000}), \eprint{hep-lat/0002020}.

\bibitem[{\citenamefont{Bonnet et~al.}(2001)\citenamefont{Bonnet, Bowman,
  Leinweber, Williams, and Zanotti}}]{Bonnet:2001uh}
\bibinfo{author}{\bibfnamefont{F.~D.~R.} \bibnamefont{Bonnet}},
  \bibinfo{author}{\bibfnamefont{P.~O.} \bibnamefont{Bowman}},
  \bibinfo{author}{\bibfnamefont{D.~B.} \bibnamefont{Leinweber}},
  \bibinfo{author}{\bibfnamefont{A.~G.} \bibnamefont{Williams}},
  \bibnamefont{and} \bibinfo{author}{\bibfnamefont{J.~M.}
  \bibnamefont{Zanotti}}, \bibinfo{journal}{Phys. Rev.}
  \textbf{\bibinfo{volume}{D64}}, \bibinfo{pages}{034501}
  (\bibinfo{year}{2001}), \eprint{hep-lat/0101013}.

\bibitem[{\citenamefont{Bloch et~al.}(2003)\citenamefont{Bloch, Cucchieri,
  Langfeld, and Mendes}}]{Bloch:2002we}
\bibinfo{author}{\bibfnamefont{J.~C.~R.} \bibnamefont{Bloch}},
  \bibinfo{author}{\bibfnamefont{A.}~\bibnamefont{Cucchieri}},
  \bibinfo{author}{\bibfnamefont{K.}~\bibnamefont{Langfeld}}, \bibnamefont{and}
  \bibinfo{author}{\bibfnamefont{T.}~\bibnamefont{Mendes}},
  \bibinfo{journal}{Nucl. Phys. Proc. Suppl.} \textbf{\bibinfo{volume}{119}},
  \bibinfo{pages}{736} (\bibinfo{year}{2003}), \eprint{hep-lat/0209040}.

\bibitem[{\citenamefont{Bloch et~al.}(2004)\citenamefont{Bloch, Cucchieri,
  Langfeld, and Mendes}}]{Bloch:2003sk}
\bibinfo{author}{\bibfnamefont{J.~C.~R.} \bibnamefont{Bloch}},
  \bibinfo{author}{\bibfnamefont{A.}~\bibnamefont{Cucchieri}},
  \bibinfo{author}{\bibfnamefont{K.}~\bibnamefont{Langfeld}}, \bibnamefont{and}
  \bibinfo{author}{\bibfnamefont{T.}~\bibnamefont{Mendes}},
  \bibinfo{journal}{Nucl. Phys.} \textbf{\bibinfo{volume}{B687}},
  \bibinfo{pages}{76} (\bibinfo{year}{2004}), \eprint{hep-lat/0312036}.

\bibitem[{\citenamefont{Furui and Nakajima}(2004{\natexlab{a}})}]{Furui:2003jr}
\bibinfo{author}{\bibfnamefont{S.}~\bibnamefont{Furui}} \bibnamefont{and}
  \bibinfo{author}{\bibfnamefont{H.}~\bibnamefont{Nakajima}},
  \bibinfo{journal}{Phys. Rev.} \textbf{\bibinfo{volume}{D69}},
  \bibinfo{pages}{074505} (\bibinfo{year}{2004}{\natexlab{a}}),
  \eprint{hep-lat/0305010}.

\bibitem[{\citenamefont{Boucaud et~al.}(2005)\citenamefont{Boucaud, Leroy,
  Le~Yaouanc, Lokhov, Micheli, Pene, Rodriguez-Quintero, and
  Roiesnel}}]{Boucaud:2005gg}
\bibinfo{author}{\bibfnamefont{P.}~\bibnamefont{Boucaud}},
  \bibinfo{author}{\bibfnamefont{J.~P.} \bibnamefont{Leroy}},
  \bibinfo{author}{\bibfnamefont{A.}~\bibnamefont{Le~Yaouanc}},
  \bibinfo{author}{\bibfnamefont{A.~Y.} \bibnamefont{Lokhov}},
  \bibinfo{author}{\bibfnamefont{J.}~\bibnamefont{Micheli}},
  \bibinfo{author}{\bibfnamefont{O.}~\bibnamefont{Pene}},
  \bibinfo{author}{\bibfnamefont{J.}~\bibnamefont{Rodriguez-Quintero}},
  \bibnamefont{and} \bibinfo{author}{\bibfnamefont{C.}~\bibnamefont{Roiesnel}},
  \bibinfo{journal}{Phys. Rev.} \textbf{\bibinfo{volume}{D72}},
  \bibinfo{pages}{114503} (\bibinfo{year}{2005}), \eprint{hep-lat/0506031}.

\bibitem[{\citenamefont{Sternbeck et~al.}(2005)\citenamefont{Sternbeck,
  Ilgenfritz, M{\"u}ller-Preussker, and Schiller}}]{Sternbeck:2005tk}
\bibinfo{author}{\bibfnamefont{A.}~\bibnamefont{Sternbeck}},
  \bibinfo{author}{\bibfnamefont{E.-M.} \bibnamefont{Ilgenfritz}},
  \bibinfo{author}{\bibfnamefont{M.}~\bibnamefont{M{\"u}ller-Preussker}},
  \bibnamefont{and} \bibinfo{author}{\bibfnamefont{A.}~\bibnamefont{Schiller}},
  \bibinfo{journal}{Phys. Rev.} \textbf{\bibinfo{volume}{D72}},
  \bibinfo{pages}{014507} (\bibinfo{year}{2005}), \eprint{hep-lat/0506007}.

\bibitem[{\citenamefont{Bowman et~al.}(2007)\citenamefont{Bowman, Heller,
  Leinweber, Parappilly, Sternbeck, von Smekal, Williams, and
  Zhang}}]{Bowman:2007du}
\bibinfo{author}{\bibfnamefont{P.~O.} \bibnamefont{Bowman}},
  \bibinfo{author}{\bibfnamefont{U.~M.} \bibnamefont{Heller}},
  \bibinfo{author}{\bibfnamefont{D.~B.} \bibnamefont{Leinweber}},
  \bibinfo{author}{\bibfnamefont{M.~B.} \bibnamefont{Parappilly}},
  \bibinfo{author}{\bibfnamefont{A.}~\bibnamefont{Sternbeck}},
  \bibinfo{author}{\bibfnamefont{L.}~\bibnamefont{von Smekal}},
  \bibinfo{author}{\bibfnamefont{A.~G.} \bibnamefont{Williams}},
  \bibnamefont{and} \bibinfo{author}{\bibfnamefont{J.}~\bibnamefont{Zhang}},
  \bibinfo{journal}{Phys. Rev.} \textbf{\bibinfo{volume}{D76}},
  \bibinfo{pages}{094505} (\bibinfo{year}{2007}), \eprint{hep-lat/0703022}.

\bibitem[{\citenamefont{Cucchieri and Mendes}(2007)}]{Cucchieri:2007md}
\bibinfo{author}{\bibfnamefont{A.}~\bibnamefont{Cucchieri}} \bibnamefont{and}
  \bibinfo{author}{\bibfnamefont{T.}~\bibnamefont{Mendes}},
  \bibinfo{journal}{PoS} \textbf{\bibinfo{volume}{LAT2007}},
  \bibinfo{pages}{297} (\bibinfo{year}{2007}), \eprint{0710.0412}.

\bibitem[{\citenamefont{Cucchieri and
  Mendes}(2008{\natexlab{a}})}]{Cucchieri:2007rg}
\bibinfo{author}{\bibfnamefont{A.}~\bibnamefont{Cucchieri}} \bibnamefont{and}
  \bibinfo{author}{\bibfnamefont{T.}~\bibnamefont{Mendes}},
  \bibinfo{journal}{Phys.Rev.Lett.} \textbf{\bibinfo{volume}{100}},
  \bibinfo{pages}{241601} (\bibinfo{year}{2008}{\natexlab{a}}),
  \eprint{0712.3517}.

\bibitem[{\citenamefont{Cucchieri et~al.}(2007)\citenamefont{Cucchieri, Mendes,
  Oliveira, and Silva}}]{Cucchieri:2007zm}
\bibinfo{author}{\bibfnamefont{A.}~\bibnamefont{Cucchieri}},
  \bibinfo{author}{\bibfnamefont{T.}~\bibnamefont{Mendes}},
  \bibinfo{author}{\bibfnamefont{O.}~\bibnamefont{Oliveira}}, \bibnamefont{and}
  \bibinfo{author}{\bibfnamefont{P.}~\bibnamefont{Silva}},
  \bibinfo{journal}{Phys.Rev.} \textbf{\bibinfo{volume}{D76}},
  \bibinfo{pages}{114507} (\bibinfo{year}{2007}), \eprint{0705.3367}.

\bibitem[{\citenamefont{Sternbeck et~al.}(2007)\citenamefont{Sternbeck, von
  Smekal, Leinweber, and Williams}}]{Sternbeck:2007ug}
\bibinfo{author}{\bibfnamefont{A.}~\bibnamefont{Sternbeck}},
  \bibinfo{author}{\bibfnamefont{L.}~\bibnamefont{von Smekal}},
  \bibinfo{author}{\bibfnamefont{D.~B.} \bibnamefont{Leinweber}},
  \bibnamefont{and} \bibinfo{author}{\bibfnamefont{A.~G.}
  \bibnamefont{Williams}}, \bibinfo{journal}{PoS}
  \textbf{\bibinfo{volume}{LAT2007}}, \bibinfo{pages}{340}
  (\bibinfo{year}{2007}), \eprint{0710.1982}.

\bibitem[{\citenamefont{Cucchieri and
  Mendes}(2008{\natexlab{b}})}]{Cucchieri:2008fc}
\bibinfo{author}{\bibfnamefont{A.}~\bibnamefont{Cucchieri}} \bibnamefont{and}
  \bibinfo{author}{\bibfnamefont{T.}~\bibnamefont{Mendes}},
  \bibinfo{journal}{Phys.Rev.} \textbf{\bibinfo{volume}{D78}},
  \bibinfo{pages}{094503} (\bibinfo{year}{2008}{\natexlab{b}}),
  \eprint{0804.2371}.

\bibitem[{\citenamefont{Oliveira and Silva}(2009)}]{Oliveira:2008uf}
\bibinfo{author}{\bibfnamefont{O.}~\bibnamefont{Oliveira}} \bibnamefont{and}
  \bibinfo{author}{\bibfnamefont{P.~J.} \bibnamefont{Silva}},
  \bibinfo{journal}{Phys. Rev.} \textbf{\bibinfo{volume}{D79}},
  \bibinfo{pages}{031501} (\bibinfo{year}{2009}), \eprint{0809.0258}.

\bibitem[{\citenamefont{Bogolubsky et~al.}(2009)\citenamefont{Bogolubsky,
  Ilgenfritz, M{\"u}ller-Preussker, and Sternbeck}}]{Bogolubsky:2009dc}
\bibinfo{author}{\bibfnamefont{I.~L.} \bibnamefont{Bogolubsky}},
  \bibinfo{author}{\bibfnamefont{E.-M.} \bibnamefont{Ilgenfritz}},
  \bibinfo{author}{\bibfnamefont{M.}~\bibnamefont{M{\"u}ller-Preussker}},
  \bibnamefont{and}
  \bibinfo{author}{\bibfnamefont{A.}~\bibnamefont{Sternbeck}},
  \bibinfo{journal}{Phys. Lett.} \textbf{\bibinfo{volume}{B676}},
  \bibinfo{pages}{69} (\bibinfo{year}{2009}), \eprint{0901.0736}.

\bibitem[{\citenamefont{Oliveira and Silva}(2012)}]{Oliveira:2012eh}
\bibinfo{author}{\bibfnamefont{O.}~\bibnamefont{Oliveira}} \bibnamefont{and}
  \bibinfo{author}{\bibfnamefont{P.~J.} \bibnamefont{Silva}},
  \bibinfo{journal}{Phys. Rev.} \textbf{\bibinfo{volume}{D86}},
  \bibinfo{pages}{114513} (\bibinfo{year}{2012}), \eprint{1207.3029}.

\bibitem[{\citenamefont{Maas}(2015)}]{Maas:2014xma}
\bibinfo{author}{\bibfnamefont{A.}~\bibnamefont{Maas}},
  \bibinfo{journal}{Phys.Rev.} \textbf{\bibinfo{volume}{D91}},
  \bibinfo{pages}{034502} (\bibinfo{year}{2015}), \eprint{1402.5050}.

\bibitem[{\citenamefont{Cucchieri}(1997)}]{Cucchieri:1997dx}
\bibinfo{author}{\bibfnamefont{A.}~\bibnamefont{Cucchieri}},
  \bibinfo{journal}{Nucl. Phys.} \textbf{\bibinfo{volume}{B508}},
  \bibinfo{pages}{353} (\bibinfo{year}{1997}), \eprint{hep-lat/9705005}.

\bibitem[{\citenamefont{Bakeev et~al.}(2004)\citenamefont{Bakeev, Ilgenfritz,
  Mitrjushkin, and M{\"u}ller-Preussker}}]{Bakeev:2003rr}
\bibinfo{author}{\bibfnamefont{T.~D.} \bibnamefont{Bakeev}},
  \bibinfo{author}{\bibfnamefont{E.-M.} \bibnamefont{Ilgenfritz}},
  \bibinfo{author}{\bibfnamefont{V.~K.} \bibnamefont{Mitrjushkin}},
  \bibnamefont{and}
  \bibinfo{author}{\bibfnamefont{M.}~\bibnamefont{M{\"u}ller-Preussker}},
  \bibinfo{journal}{Phys. Rev.} \textbf{\bibinfo{volume}{D69}},
  \bibinfo{pages}{074507} (\bibinfo{year}{2004}), \eprint{hep-lat/0311041}.

\bibitem[{\citenamefont{Furui and Nakajima}(2004{\natexlab{b}})}]{Furui:2003mz}
\bibinfo{author}{\bibfnamefont{S.}~\bibnamefont{Furui}} \bibnamefont{and}
  \bibinfo{author}{\bibfnamefont{H.}~\bibnamefont{Nakajima}},
  \bibinfo{journal}{AIP Conf. Proc.} \textbf{\bibinfo{volume}{717}},
  \bibinfo{pages}{685} (\bibinfo{year}{2004}{\natexlab{b}}),
  \bibinfo{note}{[,685(2003)]}, \eprint{hep-lat/0309166}.

\bibitem[{\citenamefont{Silva and Oliveira}(2004)}]{Silva:2004bv}
\bibinfo{author}{\bibfnamefont{P.~J.} \bibnamefont{Silva}} \bibnamefont{and}
  \bibinfo{author}{\bibfnamefont{O.}~\bibnamefont{Oliveira}},
  \bibinfo{journal}{Nucl. Phys.} \textbf{\bibinfo{volume}{B690}},
  \bibinfo{pages}{177} (\bibinfo{year}{2004}), \eprint{hep-lat/0403026}.

\bibitem[{\citenamefont{Furui and Nakajima}(2004{\natexlab{c}})}]{Furui:2004cx}
\bibinfo{author}{\bibfnamefont{S.}~\bibnamefont{Furui}} \bibnamefont{and}
  \bibinfo{author}{\bibfnamefont{H.}~\bibnamefont{Nakajima}},
  \bibinfo{journal}{Phys. Rev.} \textbf{\bibinfo{volume}{D70}},
  \bibinfo{pages}{094504} (\bibinfo{year}{2004}{\natexlab{c}}).

\bibitem[{\citenamefont{Bogolubsky et~al.}(2006)\citenamefont{Bogolubsky,
  Burgio, Mitrjushkin, and M{\"u}ller-Preussker}}]{Bogolubsky:2005wf}
\bibinfo{author}{\bibfnamefont{I.~L.} \bibnamefont{Bogolubsky}},
  \bibinfo{author}{\bibfnamefont{G.}~\bibnamefont{Burgio}},
  \bibinfo{author}{\bibfnamefont{V.~K.} \bibnamefont{Mitrjushkin}},
  \bibnamefont{and}
  \bibinfo{author}{\bibfnamefont{M.}~\bibnamefont{M{\"u}ller-Preussker}},
  \bibinfo{journal}{Phys. Rev.} \textbf{\bibinfo{volume}{D74}},
  \bibinfo{pages}{034503} (\bibinfo{year}{2006}), \eprint{hep-lat/0511056}.

\bibitem[{\citenamefont{Bornyakov et~al.}(2009)\citenamefont{Bornyakov,
  Mitrjushkin, and M{\"u}ller-Preussker}}]{Bornyakov:2008yx}
\bibinfo{author}{\bibfnamefont{V.~G.} \bibnamefont{Bornyakov}},
  \bibinfo{author}{\bibfnamefont{V.~K.} \bibnamefont{Mitrjushkin}},
  \bibnamefont{and}
  \bibinfo{author}{\bibfnamefont{M.}~\bibnamefont{M{\"u}ller-Preussker}},
  \bibinfo{journal}{Phys. Rev.} \textbf{\bibinfo{volume}{D79}},
  \bibinfo{pages}{074504} (\bibinfo{year}{2009}), \eprint{0812.2761}.

\bibitem[{\citenamefont{Maas}(2009)}]{Maas:2008ri}
\bibinfo{author}{\bibfnamefont{A.}~\bibnamefont{Maas}}, \bibinfo{journal}{Phys.
  Rev.} \textbf{\bibinfo{volume}{D79}}, \bibinfo{pages}{014505}
  (\bibinfo{year}{2009}), \eprint{0808.3047}.

\bibitem[{\citenamefont{Maas et~al.}(2010)\citenamefont{Maas, Pawlowski,
  Spielmann, Sternbeck, and von Smekal}}]{Maas:2009ph}
\bibinfo{author}{\bibfnamefont{A.}~\bibnamefont{Maas}},
  \bibinfo{author}{\bibfnamefont{J.~M.} \bibnamefont{Pawlowski}},
  \bibinfo{author}{\bibfnamefont{D.}~\bibnamefont{Spielmann}},
  \bibinfo{author}{\bibfnamefont{A.}~\bibnamefont{Sternbeck}},
  \bibnamefont{and} \bibinfo{author}{\bibfnamefont{L.}~\bibnamefont{von
  Smekal}}, \bibinfo{journal}{Eur. Phys. J.} \textbf{\bibinfo{volume}{C68}},
  \bibinfo{pages}{183} (\bibinfo{year}{2010}), \eprint{0912.4203}.

\bibitem[{\citenamefont{Hughes et~al.}(2013)\citenamefont{Hughes, Mehta, and
  Skullerud}}]{Hughes:2012hg}
\bibinfo{author}{\bibfnamefont{C.}~\bibnamefont{Hughes}},
  \bibinfo{author}{\bibfnamefont{D.}~\bibnamefont{Mehta}}, \bibnamefont{and}
  \bibinfo{author}{\bibfnamefont{J.-I.} \bibnamefont{Skullerud}},
  \bibinfo{journal}{Annals Phys.} \textbf{\bibinfo{volume}{331}},
  \bibinfo{pages}{188} (\bibinfo{year}{2013}), \eprint{1203.4847}.

\bibitem[{\citenamefont{Maas}(2010)}]{Maas:2009se}
\bibinfo{author}{\bibfnamefont{A.}~\bibnamefont{Maas}}, \bibinfo{journal}{Phys.
  Lett.} \textbf{\bibinfo{volume}{B689}}, \bibinfo{pages}{107}
  (\bibinfo{year}{2010}), \eprint{0907.5185}.

\bibitem[{\citenamefont{Sternbeck and
  M{\"u}ller-Preussker}(2013)}]{Sternbeck:2012mf}
\bibinfo{author}{\bibfnamefont{A.}~\bibnamefont{Sternbeck}} \bibnamefont{and}
  \bibinfo{author}{\bibfnamefont{M.}~\bibnamefont{M{\"u}ller-Preussker}},
  \bibinfo{journal}{Phys.Lett.} \textbf{\bibinfo{volume}{B726}},
  \bibinfo{pages}{396} (\bibinfo{year}{2013}), \eprint{1211.3057}.

\bibitem[{\citenamefont{Zwanziger}(2013)}]{Zwanziger:2012xg}
\bibinfo{author}{\bibfnamefont{D.}~\bibnamefont{Zwanziger}},
  \bibinfo{journal}{Phys.Rev.} \textbf{\bibinfo{volume}{D87}},
  \bibinfo{pages}{085039} (\bibinfo{year}{2013}), \eprint{1209.1974}.

\bibitem[{\citenamefont{Cucchieri et~al.}(2012)\citenamefont{Cucchieri, Dudal,
  and Vandersickel}}]{Cucchieri:2012cb}
\bibinfo{author}{\bibfnamefont{A.}~\bibnamefont{Cucchieri}},
  \bibinfo{author}{\bibfnamefont{D.}~\bibnamefont{Dudal}}, \bibnamefont{and}
  \bibinfo{author}{\bibfnamefont{N.}~\bibnamefont{Vandersickel}},
  \bibinfo{journal}{Phys.Rev.} \textbf{\bibinfo{volume}{D85}},
  \bibinfo{pages}{085025} (\bibinfo{year}{2012}), \eprint{1202.1912}.

\bibitem[{\citenamefont{von Smekal et~al.}(2008)\citenamefont{von Smekal,
  Jorkowski, Mehta, and Sternbeck}}]{vonSmekal:2008es}
\bibinfo{author}{\bibfnamefont{L.}~\bibnamefont{von Smekal}},
  \bibinfo{author}{\bibfnamefont{A.}~\bibnamefont{Jorkowski}},
  \bibinfo{author}{\bibfnamefont{D.}~\bibnamefont{Mehta}}, \bibnamefont{and}
  \bibinfo{author}{\bibfnamefont{A.}~\bibnamefont{Sternbeck}},
  \bibinfo{journal}{PoS} \textbf{\bibinfo{volume}{CONFINEMENT8}},
  \bibinfo{pages}{048} (\bibinfo{year}{2008}), \eprint{0812.2992}.

\bibitem[{\citenamefont{Burgio et~al.}(2010)\citenamefont{Burgio, Quandt, and
  Reinhardt}}]{Burgio:2009xp}
\bibinfo{author}{\bibfnamefont{G.}~\bibnamefont{Burgio}},
  \bibinfo{author}{\bibfnamefont{M.}~\bibnamefont{Quandt}}, \bibnamefont{and}
  \bibinfo{author}{\bibfnamefont{H.}~\bibnamefont{Reinhardt}},
  \bibinfo{journal}{Phys. Rev.} \textbf{\bibinfo{volume}{D81}},
  \bibinfo{pages}{074502} (\bibinfo{year}{2010}), \eprint{0911.5101}.

\bibitem[{\citenamefont{Cucchieri
  et~al.}(2014{\natexlab{a}})\citenamefont{Cucchieri, Dudal, Mendes, and
  Vandersickel}}]{Cucchieri:2014via}
\bibinfo{author}{\bibfnamefont{A.}~\bibnamefont{Cucchieri}},
  \bibinfo{author}{\bibfnamefont{D.}~\bibnamefont{Dudal}},
  \bibinfo{author}{\bibfnamefont{T.}~\bibnamefont{Mendes}}, \bibnamefont{and}
  \bibinfo{author}{\bibfnamefont{N.}~\bibnamefont{Vandersickel}},
  \bibinfo{journal}{Phys.Rev.} \textbf{\bibinfo{volume}{D90}},
  \bibinfo{pages}{051501} (\bibinfo{year}{2014}{\natexlab{a}}),
  \eprint{1405.1547}.

\bibitem[{\citenamefont{Cucchieri
  et~al.}(2014{\natexlab{b}})\citenamefont{Cucchieri, Dudal, Mendes, and
  Vandersickel}}]{Cucchieri:2014xfa}
\bibinfo{author}{\bibfnamefont{A.}~\bibnamefont{Cucchieri}},
  \bibinfo{author}{\bibfnamefont{D.}~\bibnamefont{Dudal}},
  \bibinfo{author}{\bibfnamefont{T.}~\bibnamefont{Mendes}}, \bibnamefont{and}
  \bibinfo{author}{\bibfnamefont{N.}~\bibnamefont{Vandersickel}},
  \bibinfo{journal}{PoS} \textbf{\bibinfo{volume}{LATTICE2014}},
  \bibinfo{pages}{347} (\bibinfo{year}{2014}{\natexlab{b}}),
  \eprint{1410.8410}.

\bibitem[{\citenamefont{Bornyakov et~al.}(2010)\citenamefont{Bornyakov,
  Mitrjushkin, and M{\"u}ller-Preussker}}]{Bornyakov:2009ug}
\bibinfo{author}{\bibfnamefont{V.~G.} \bibnamefont{Bornyakov}},
  \bibinfo{author}{\bibfnamefont{V.~K.} \bibnamefont{Mitrjushkin}},
  \bibnamefont{and}
  \bibinfo{author}{\bibfnamefont{M.}~\bibnamefont{M{\"u}ller-Preussker}},
  \bibinfo{journal}{Phys. Rev.} \textbf{\bibinfo{volume}{D81}},
  \bibinfo{pages}{054503} (\bibinfo{year}{2010}), \eprint{0912.4475}.

\bibitem[{\citenamefont{von Smekal et~al.}(2009)\citenamefont{von Smekal,
  Maltman, and Sternbeck}}]{vonSmekal:2009ae}
\bibinfo{author}{\bibfnamefont{L.}~\bibnamefont{von Smekal}},
  \bibinfo{author}{\bibfnamefont{K.}~\bibnamefont{Maltman}}, \bibnamefont{and}
  \bibinfo{author}{\bibfnamefont{A.}~\bibnamefont{Sternbeck}},
  \bibinfo{journal}{Phys. Lett.} \textbf{\bibinfo{volume}{B681}},
  \bibinfo{pages}{336} (\bibinfo{year}{2009}), \eprint{0903.1696}.

\bibitem[{\citenamefont{Zwanziger}(1994)}]{Zwanziger:1993dh}
\bibinfo{author}{\bibfnamefont{D.}~\bibnamefont{Zwanziger}},
  \bibinfo{journal}{Nucl. Phys.} \textbf{\bibinfo{volume}{B412}},
  \bibinfo{pages}{657} (\bibinfo{year}{1994}).

\bibitem[{\citenamefont{Fingberg et~al.}(1993)\citenamefont{Fingberg, Heller,
  and Karsch}}]{Fingberg:1992ju}
\bibinfo{author}{\bibfnamefont{J.}~\bibnamefont{Fingberg}},
  \bibinfo{author}{\bibfnamefont{U.~M.} \bibnamefont{Heller}},
  \bibnamefont{and} \bibinfo{author}{\bibfnamefont{F.}~\bibnamefont{Karsch}},
  \bibinfo{journal}{Nucl. Phys.} \textbf{\bibinfo{volume}{B392}},
  \bibinfo{pages}{493} (\bibinfo{year}{1993}), \eprint{hep-lat/9208012}.

\bibitem[{\citenamefont{Lucini and Teper}(2001)}]{Lucini:2001ej}
\bibinfo{author}{\bibfnamefont{B.}~\bibnamefont{Lucini}} \bibnamefont{and}
  \bibinfo{author}{\bibfnamefont{M.}~\bibnamefont{Teper}},
  \bibinfo{journal}{JHEP} \textbf{\bibinfo{volume}{0106}}, \bibinfo{pages}{050}
  (\bibinfo{year}{2001}), \eprint{hep-lat/0103027}.

\bibitem[{\citenamefont{Bogolubsky et~al.}(2008)\citenamefont{Bogolubsky,
  Bornyakov, Burgio, Ilgenfritz, Mitrjushkin, and
  M{\"u}ller-Preussker}}]{Bogolubsky:2007bw}
\bibinfo{author}{\bibfnamefont{I.~L.} \bibnamefont{Bogolubsky}},
  \bibinfo{author}{\bibfnamefont{V.~G.} \bibnamefont{Bornyakov}},
  \bibinfo{author}{\bibfnamefont{G.}~\bibnamefont{Burgio}},
  \bibinfo{author}{\bibfnamefont{E.-M.} \bibnamefont{Ilgenfritz}},
  \bibinfo{author}{\bibfnamefont{V.~K.} \bibnamefont{Mitrjushkin}},
  \bibnamefont{and}
  \bibinfo{author}{\bibfnamefont{M.}~\bibnamefont{M{\"u}ller-Preussker}},
  \bibinfo{journal}{Phys. Rev.} \textbf{\bibinfo{volume}{D77}},
  \bibinfo{pages}{014504} (\bibinfo{year}{2008}), \eprint{0707.3611}.

\bibitem[{\citenamefont{Schemel}(2006)}]{Schemel:2006da}
\bibinfo{author}{\bibfnamefont{P.}~\bibnamefont{Schemel}},
  \bibinfo{type}{Diploma thesis}, \bibinfo{school}{Humboldt University Berlin,
  Germany, see http://pha.physik.hu-berlin.de - Theses} (\bibinfo{year}{2006}).

\bibitem[{\citenamefont{Bogolubsky et~al.}(2007)\citenamefont{Bogolubsky,
  Bornyakov, Burgio, Ilgenfritz, Mitrjushkin, M{\"u}ller-Preussker, and
  Schemel}}]{Bogolubsky:2007pq}
\bibinfo{author}{\bibfnamefont{I.~L.} \bibnamefont{Bogolubsky}},
  \bibinfo{author}{\bibfnamefont{V.~G.} \bibnamefont{Bornyakov}},
  \bibinfo{author}{\bibfnamefont{G.}~\bibnamefont{Burgio}},
  \bibinfo{author}{\bibfnamefont{E.-M.} \bibnamefont{Ilgenfritz}},
  \bibinfo{author}{\bibfnamefont{V.~K.} \bibnamefont{Mitrjushkin}},
  \bibinfo{author}{\bibfnamefont{M.}~\bibnamefont{M{\"u}ller-Preussker}},
  \bibnamefont{and} \bibinfo{author}{\bibfnamefont{P.}~\bibnamefont{Schemel}},
  \bibinfo{journal}{PoS} \textbf{\bibinfo{volume}{LAT2007}},
  \bibinfo{pages}{318} (\bibinfo{year}{2007}), \eprint{0710.3234}.

\bibitem[{\citenamefont{Nakagawa et~al.}(2009)\citenamefont{Nakagawa, Voigt,
  Ilgenfritz, M{\"u}ller-Preussker, Nakamura, Saito, Sternbeck, and
  Toki}}]{Nakagawa:2009zf}
\bibinfo{author}{\bibfnamefont{Y.}~\bibnamefont{Nakagawa}},
  \bibinfo{author}{\bibfnamefont{A.}~\bibnamefont{Voigt}},
  \bibinfo{author}{\bibfnamefont{E.-M.} \bibnamefont{Ilgenfritz}},
  \bibinfo{author}{\bibfnamefont{M.}~\bibnamefont{M{\"u}ller-Preussker}},
  \bibinfo{author}{\bibfnamefont{A.}~\bibnamefont{Nakamura}},
  \bibinfo{author}{\bibfnamefont{T.}~\bibnamefont{Saito}},
  \bibinfo{author}{\bibfnamefont{A.}~\bibnamefont{Sternbeck}},
  \bibnamefont{and} \bibinfo{author}{\bibfnamefont{H.}~\bibnamefont{Toki}},
  \bibinfo{journal}{Phys. Rev.} \textbf{\bibinfo{volume}{D79}},
  \bibinfo{pages}{114504} (\bibinfo{year}{2009}), \eprint{0902.4321}.

\bibitem[{\citenamefont{Boucaud et~al.}(2012)\citenamefont{Boucaud, Leroy,
  Yaouanc, Micheli, Pene, and Rodriguez-Quintero}}]{Boucaud:2011ug}
\bibinfo{author}{\bibfnamefont{P.}~\bibnamefont{Boucaud}},
  \bibinfo{author}{\bibfnamefont{J.~P.} \bibnamefont{Leroy}},
  \bibinfo{author}{\bibfnamefont{A.~L.} \bibnamefont{Yaouanc}},
  \bibinfo{author}{\bibfnamefont{J.}~\bibnamefont{Micheli}},
  \bibinfo{author}{\bibfnamefont{O.}~\bibnamefont{Pene}}, \bibnamefont{and}
  \bibinfo{author}{\bibfnamefont{J.}~\bibnamefont{Rodriguez-Quintero}},
  \bibinfo{journal}{Few Body Syst.} \textbf{\bibinfo{volume}{53}},
  \bibinfo{pages}{387} (\bibinfo{year}{2012}), \eprint{1109.1936}.

\end{thebibliography}

\end{document}